\documentclass[12pt]{jhep3}
\usepackage{graphicx}
\usepackage{amsmath}\usepackage{amssymb}

\makeatletter
\newdimen\rh@wd
\newdimen\rh@hta
\newdimen\rh@htb
\newbox\rh@box
\def\rh@measure#1{\setbox\rh@box=\hbox{$#1$}\rh@wd=\wd\rh@box
\rh@hta=\ht\rh@box}

\def\widecheck#1{\rh@measure{#1}%
   \setbox\rh@box=\hbox{$\widehat{\vrule height \rh@hta width\z@
\kern\rh@wd}$}%
   \rh@htb=\ht\rh@box \advance\rh@htb\rh@hta \advance\rh@htb\p@
   \ooalign{$\vrule height \ht\rh@box width\z@ #1$\cr
            \raise\rh@htb\hbox{\scalebox{1}[-1]{\box\rh@box}}\cr}}
\makeatother

\title{Infinite Dimensional Symmetries of Self Dual Yang Mills} 
\author{ Paul Mansfield and Adam Wardlow  \\
Department of Mathematical Sciences, University of Durham\\
South Road, Durham, DH1 3LE, U.K.\\ 
E-mails:
 \email{p.r.w.mansfield@durham.ac.uk}, \email{a.b.wardlow@durham.ac.uk}
}
\abstract{We construct symmetries of the Chalmers-Siegel action describing self-dual Yang-Mills theory using a canonical transformation to a free theory. The symmetries form an infinite dimensional Lie algebra in the group algebra of isometries.}
\keywords{Discrete and Finite Symmetries, Space-Time Symmetries, QCD}
\preprint{DCPT-09/27}
\begin{document}
\numberwithin{equation}{section}
\maketitle
\section{Introduction}

The observation that tree-level gluon scattering amplitudes localise on simple curves in twistor space \cite{witten-2004-252} led to the proposal of a new set of rules for calculating such amplitudes \cite{cachazo-2004-0409}. These provided an efficient alternative to conventional Feynman rules. Initially they were proven using non-Lagrangian methods \cite{britto-2005-94}, but they may be derived by applying a non-local canonical transformation to light-cone
Yang-Mills theory \cite{Gorsky:2005sf}, \cite{Pauls_paper}. This action can be split into a part, the Chalmers-Siegel action, \cite{Chalmers:1996rq}, that describes self-dual gauge theory  and the rest. By itself the self-dual theory has the bizarre property of yielding an S-matrix that is trivial at tree-level whilst having non-linear Euler-Lagrange equations, and
non-trivial scattering amplitudes at one-loop. The canonical transformation maps the Chalmers-Siegel part of the Lagrangian to a free theory, so that the rest of the Lagrangian furnishes interaction terms. This canonical transformation provides a new approach to the self-dual sector of gauge theories. We will use it to construct new non-local symmetries of 
the self-dual Lagrangian, thereby extending the programme of \cite{dolan} off-shell, (see also \cite{adam}, \cite{Popov:2006qu} and \cite{Wolf:2004hp}).

As is well-known, a free-theory with Euler-Lagrange equation $\Omega(x)\,\phi (x)=0$ has a symmetry if the operator,
$\Omega$ transforms covariantly when $x\rightarrow x_G$, because if  $\Omega(x_G)=A\,\Omega(x)$,
then $0=\Omega(x_G)\,\phi (x_G)=A\,\Omega(x)\,\phi (x_G)$, so $\phi (x_G)$ is a new solution.
Taking the transformation $G$ close to the identity gives the change in the field,
$\delta\,\phi(x)=\phi (x_G)-\phi(x)$ which
can be used to construct the usual Noether currents and conserved charges. However, because the 
Euler-Lagrange equation is linear we can also construct a new solution as $\phi (x)+
\epsilon\,\phi (x_G)$, with $G$ a finite transformation. The change in the field is then
$\delta\,\phi(x)=\epsilon\,\phi (x_G)$. This leads to higher derivative conserved currents such as the `zilch' of the electromagnetic field discovered in the 60s by Lipkin \cite{lipkin}. 

Since the canonical transformation maps the Chalmers-Siegel action to a free theory we can in principle construct the symmetries of this action for self-dual Yang-Mills from those of the free theory by inverting the transformation back to the original variables. This leads to quite cumbersome expressions, so to produce a compact result we begin by examining just the first few orders (in powers of the fields) of the transformations $A\rightarrow A+\delta A$ and $\overline{A}\rightarrow \overline{A}+\delta\overline{A}$ and guess a more concise general expression for $\delta A$ given by
\begin{eqnarray}
	\nonumber\delta A_1=-\epsilon\sum_{n=2}^{\infty}\sum_{i=2}^{n}\sum_{j=i}^n\int_{2\cdots n}\frac{\hat{1}}{\hat{q}}\Gamma(q^G,i^G,\cdots,j^G)\Gamma(q,j+1,\cdots,n,1\cdots,i-1)\times\\
\nonumber	\times A_{\bar{2}}\cdots A_{\bar{i}^G}\cdots A_{\bar{j}^G}\cdots A_{\bar{n}}
\end{eqnarray} 
and expanded diagramatically in fig (3), which also includes the expansion for $\delta\overline{A}$. We then prove that this guess is correct by showing that it leaves the Chalmers-Siegel action invariant.

\section{Review of the Lagrangian Formulation of MHV Rules}
 In recent years, an alternative approach to the usual Feynman diagram expansion of Yang-Mills theory has been suggested at tree level, \cite{cachazo-2004-0409}, and to low order in the loop expansion. The Feynman approach is well understood but the complexity of the calculations grows very quickly. In many cases scattering amplitudes are much simpler than their constituent Feynman diagrams. For example
the Parke-Taylor amplitude \cite{Parke:1986gb} for a tree-level process in which the greatest number of gluon helicities changes is written in terms of the reduced amplitude

\begin{equation}
\nonumber	A=g^{n-2}\frac{\left\langle \lambda_r,\lambda_s\right\rangle^4}{\prod^{n}_{j=1}\left\langle \lambda_j,\lambda_{j+1}\right\rangle}
\end{equation}
where g is the coupling constant and r and s label the gluons with positive and negative helicity respectively. The $\lambda_j$ are two spinors satisfying 
\begin{equation}
\nonumber	\lambda_j \widetilde{\lambda}_j=p^t\,{1}+\Sigma \sigma^i p^i
\end{equation}
with $\sigma^i$ being the Pauli matrices and $p^i$ being the momenta of the on-shell gluons. The bracket $\left\langle\ \ ,\ \  \right\rangle$  is $\left\langle \lambda_j, \lambda_k\right\rangle=\lambda^T_j i \sigma^2 \lambda_k$. Then, the full tree level amplitude is a sum over colour ordered amplitudes:
\begin{equation}
\nonumber	\mathcal{A}_n=\sum_\sigma tr\left(T^{R_\sigma(1)}\cdots T^{R_\sigma(n)}\right)i(2\pi)^4 \delta^4(p^1+\cdots+p^n)\,A^{\sigma}_{n}.
\end{equation}
These amplitudes (suitably continued off-shell)  become the interaction vertices of the CSW approach to Yang-Mills \cite{cachazo-2004-0409} and \cite{witten-2004-252}.
These MHV rules were proven outside the Lagrangian formalism, indirectly from the BCFW recursion \cite{britto-2005-94} and using twistor methods.  (See \cite{boels-2007-014} through to \nocite{mason-2006-636,boels-2007-648,boels-2007-76,boels-2007-016,boels-2008-007,boels-2008-183} \cite{Wen}.) An alternative, Lagrangian approach was taken in \cite{Gorsky:2005sf} and \cite{Pauls_paper} which describe a canonical transformation taking the standard Yang-Mills action into one generating the MHV rules. See also \cite{ettle-2007-011} and \cite{Fu:2009nh}. We shall now give a brief review.

The Yang-Mills action in coordinates $(t,x^1,x^2,x^3)$ is

\begin{equation}
\nonumber	S=\frac{1}{2g^2}\int{dt dx^1 dx^2 dx^3}tr\left(F^{\mu\nu}F_{\mu\nu}\right)
\end{equation}
where the trace is taken over the generators of the gauge group $T^R$, and
\begin{align}
	\nonumber F_{\mu\nu}&=\left[D_{\mu},D_{\nu}\right]  &   D_{\mu}&=\partial_{\mu}+A_{\mu}\\
	\nonumber A_{\mu}&=A_{\mu}^R T^R   &    \left[T^R,T^S\right]&=f^{RSP}T^P\\
	\nonumber tr(T^R T^S)&=-\frac{\delta^{RS}}{2}.
\end{align}
We will use light-front co-ordinates
$x^0=t-x^3$,  $x^{\overline{0}}=t+x^3$, $z=x^1+ix^2$ and $\overline{z}=x^1-ix^2$.
By imposing the gauge condition $A_{\overline{0}}=0$, and integrating out the non-dynamical field $A_{0}$  we arrive at the transformed action
\begin{equation}
	\label{eq:fulllightconeaction}
	S=\frac{4}{g^2} \int dx^0 \left\{L^{-+}+L^{++-}+L^{--+}+L^{--++}\right\}
\end{equation}
where the L's are the terms in the lagrangian, which is defined on the light front surface as an integral over constant $x^0$ surfaces. The decorations on the L's label the helicity content and we observe that the term $L^{++-}$ is unwanted since it contains only one negative helicity, whereas we need two negative helicities in the MHV formalism. Further, the terms $L^{++...++--}$ are missing. On the quantization surface, it is worth noting that the fields have the same $x^0$ dependence so we don't have to explicitly write this and we use the notation $(x^{\overline{0}},z,\overline{z})=\textbf{x}$ on the quantisation surface. Explicitly, the L's are given by \cite{Pauls_paper}
\begin{eqnarray}
\nonumber	L^{+-}[A] &=& \frac{4}{g^2}tr\int_{\Sigma} d^3\textbf{x} \overline{A}\left(\partial_0\partial_{\overline{0}}-\partial_z\partial_{\overline{z}}\right)A\\
\nonumber	L^{++-}[A] &=& \frac{4}{g^2}tr\int_{\Sigma} d^3\textbf{x}\left(-\partial_{\overline{z}}\partial_{\overline{0}}^{-1}A\right)\left[A , \partial_{\overline{0}}\overline{A}\right]\\
\nonumber	L^{--+}[A] &=&\frac{4}{g^2}tr\int_{\Sigma}d^3 \overline{x}\left[\overline{A},\partial_{\overline{0}}A\right]\left(-\partial_z \partial_{\overline{0}}^{-1}\overline{A}\right)\\
\nonumber	L^{--++}[A] &=& \frac{4}{g^2}tr\int_{\Sigma} d^3\overline{x}\left(-\left[[\overline{A},\partial_{\overline{0}}A]\partial_{\overline{0}}^{-2}\left[A,\partial_{\overline{0}}\overline{A}\right]\right]\right).
\end{eqnarray}

To remove the unwanted term $L^{++-}$ and generate the missing terms we define a change of variables $A,\,\overline{A}\rightarrow B,\,\overline B$ so that
\begin{equation}
\label{eq:field transformations}
	L^{+-}[A,\,\overline{A} ]+L^{++-}[A,\,\overline{A} ]=L^{+-}[B,\,\overline{B} ].
\end{equation}
$B$ is a functional of $A$ only on the quantization surface, $B=B[A]$, and
\begin{equation}
\label{eq:abar in terms if bbar}
	\partial_{\overline{0}}\overline{A}(\textbf{y})=\int_{\Sigma} d^3\textbf{x}\frac{\delta B(\textbf{x})}{\delta A(\textbf{y})}\partial_{\overline{0}}\overline B(\textbf{x})
\end{equation}
where $\Sigma$ refers to the quantization surface. It transpires that not only does this remove the unwanted vertex, it also generates the missing MHV vertices. The LHS of eqn (\ref{eq:field transformations}) is known as the Chalmers-Siegel action on the light cone and its Euler-Lagrange equations give the self dual Yang-Mills equations.

By substituting (\ref{eq:abar in terms if bbar}) into (\ref{eq:field transformations}) and noting that terms involving $\partial_{0}A$ and $\partial_{0}B$ are automatically equal \cite{Pauls_paper} we arrive at the defining expression relating $A$ and $B$. This is given by the following functional differential equation, (suppressing the $x^0$ dependence for brevity),
\begin{equation}
\label {eq:func diff}
	\int_\Sigma d^3\textbf{y}\left[D,\partial_{\overline{z}}\partial_{\overline{0}}^{-1}A\right]\left(\textbf{y}\right)\frac{\delta B\left(\textbf{x}\right)}{\delta A\left(\textbf{y}\right)}=\omega\left(\textbf{x}\right) B\left(\textbf{x}\right).
\end{equation}
Using this expression, one can calculate $B$ in terms of $A$, and its inverse $A$ in terms of $B$. In momentum space, (\ref{eq:func diff}) can be written
\begin{equation}
\label{eq:defining relation between A and B}
	\omega_1 A_1-i\int_{23}\left[A_2,\zeta_3 A_3 \right](2\pi)^3 \delta \left(\textbf{p}_1-\textbf{p}_2-\textbf{p}_3\right)=\int_p \omega({\textbf{p}})B(\textbf{p})\frac{\delta A(\textbf{p}_1)}{\delta B({\textbf{p}})}
\end{equation}
where we use the same notation to that in \cite{ettle_and_morris} in which $\zeta(p)=p_{\overline{z}}/p_{\overline{0}}$ and $\omega(p)=p_{z}p_{\overline{z}}/p_{\overline{0}}$. The group generators are absorbed into the fields, we introduce the notation $A_s=A(\textbf{p}_s)$, $A_{\overline{s}}=A(-\textbf{p}_s)$ and we introduce the shorthand notation
\begin{equation}
\nonumber	\int_{1 \cdots n}=\int \frac{d^4 \textbf{p}_1}{(2\pi)^4} \cdots \frac{d^4 \textbf{p}_n}{(2\pi)^4}.
\end{equation}
Ettle and Morris define the above notation as integrals over the quantisation surfaces since there is no need to Fourier transform the $x^0$ dependence. Here however, we shall be applying a linear transformation involving all the spacetime coordinates, so it makes sense to Fourier transform the $x^0$ dependence, which does not affect the calculations in \cite{ettle_and_morris}. We also introduce the notation $(p_0^n,p_{\overline{0}}^n,p_z^n,p_{\overline{z}}^n)=(\check{n},\hat{n},\tilde{n},\bar{n})$ and the following brackets, their meanings described in \cite{ettle_and_morris}
\begin{eqnarray}
\nonumber	\left\{p_1,p_2\right\}&=&\hat{1}\bar{2}-\hat{2}\bar{1}\\
\nonumber	\left(p_1,p_2\right)&=&\hat{1}\tilde{2}-\hat{2}\tilde{1}.
\end{eqnarray}
The relation (\ref{eq:defining relation between A and B}) has power series solutions of the form
\begin{equation}
\label{eq:A in terms of B}
	A_1=\sum^{\infty}_{n=2}{\int_{2\cdots n}\Upsilon\left(1 \dots n\right)B_{\overline{2}} \cdots B_{\overline{n}}}
\end{equation}
using the shorthand notation, and dropping the momentum conserving delta functions and factors of $2\pi$ (as we shall do throughout the majority of this paper). Similarly, its inverse is given by the power series
\begin{equation}
\label{eq:B in terms of A}
	B_1=\sum^{\infty}_{n=2}{\int_{2\cdots n}\Gamma\left(1 \dots n\right)A_{\overline{2}} \cdots A_{\overline{n}}}
\end{equation}

We solve for $\Gamma$ and $\Upsilon$ by putting these expressions into (\ref{eq:defining relation between A and B}) thereby extracting a recursion relation. When expressed in terms of their independent momenta, $\Upsilon(1,\cdots,n)$ and $\Gamma(1,\cdots,n)$ take the following particularly simple form
\begin{equation}
\label{eq:Upsilon}
	\Upsilon(1,\cdots,n)=(-i)^n\frac{\hat{1}}{(2,3)}\frac{\hat{3}}{(3,4)}\cdots\frac{\widehat{n-1}}{(n-1,n)}
\end{equation}
and also
\begin{equation}
\label{eq:Gamma}
\Gamma(1,\cdots,n)=-(i)^n\frac{\hat{1}}{(1,2)}\frac{\hat{1}}{(1,2+3)}\cdots\frac{\hat{1}}{(1,2+\cdots (n-1))}.
\end{equation}	
We should pay attention to the fact that these coefficients are independent of $p_0$ and $p_{\bar{z}}$ when expressed in this way.

In addition, we can express $\overline{A}$ as a power series in $\overline{B}$ viz
\begin{equation}
\label{eq:Abar in terms of Bbar}
	\overline{A}_{\bar{1}}=\sum_{n=2}^{\infty}\sum_{k=2}^{n}\int_{2\cdots n}\frac{\hat{k}}{\hat{1}}\Xi^{k}(\bar{1}2\cdots n)B_{\bar{2}}\cdots \overline{B}_{\bar{k}}\cdots B_{\bar{n}}
\end{equation}
where the coefficients are given by
\begin{equation}
\label{eq:Xi}
	\Xi^{k}(12\cdots n)=-\frac{\hat{k}}{\hat{1}}\Upsilon(12\cdots n).
\end{equation}
Note we use a different convention for the indices attached to $\Xi$. In the paper \cite{ettle_and_morris}, the left hand side of the above reads $\Xi^{k-1}$. 

Ettle and Morris \cite{ettle_and_morris} do not calculate the inverse of (\ref{eq:Abar in terms of Bbar}) but the calculation is similar to the one they describe in some respects. We begin by writing an ansatz for the inverse of (\ref{eq:Abar in terms of Bbar})
\begin{equation}
\label{eq:bhat_in_terms_of_A}
\overline{B}_{\bar{1}}=\sum_{n=2}^{\infty}\sum_{k=2}^{n}\int_{2\cdots n}\frac{\hat{k}}{\hat{1}}\Theta^{k}(\bar{1}2\cdots n)A_{\bar{2}}\cdots \overline{A}_{\bar{k}}\cdots A_{\bar{n}}.
\end{equation}
Later, we will calculate $\delta\overline{A}$ and write a transformation of the field  $\overline{A}\rightarrow\overline{A}+\delta\overline{A}$ to the first three orders in powers of the fields $A$ and $\overline{A}$. As discussed already, we will then guess a more general result to all orders and prove that it leaves the Chalmers-Siegel action invariant so it is only necessary to calculate the coefficients of first five terms in (\ref{eq:bhat_in_terms_of_A}) to use in the explicit calculations of the first three orders in $A$ and $\overline{A}$ in the expresion for $\delta\overline{A}$. We differentiate (\ref{eq:B in terms of A}) with respect to $x^{0}$, which in momentum space gives
\begin{equation}
\label{eq:diffB in terms of A}
\check{1}B_{\bar{1}}=\sum_{n=2}^{\infty}\sum_{k=2}^{n}\int_{2\cdots n}\check{k}\Gamma(\bar{1}2\cdots n)A_{\bar{2}}\cdots {A}_{\bar{k}}\cdots A_{\bar{n}}
\end{equation}
and then use
\begin{equation}
\label{eq:invariant}
tr\int_{1}\check{1}A_1\hat{\bar{1}}\overline{A}_{\bar{1}}=tr\int_{1}\check{1}B_1\hat{\bar{1}}\overline{B}_{\bar{1}}
\end{equation}
to extract a recurrence relation for $\Theta^{k}(12\cdots n)$ to the first few order in $\Theta$. By substituting eqn (\ref{eq:diffB in terms of A}) and eqn (\ref{eq:bhat_in_terms_of_A}) into the invariant quantity (\ref{eq:invariant}) and considering momentum conservation we can easily extract the first five expressions for $\Theta$,
\begin{align}
\label{eq:firstfeworders}
\nonumber\Theta^2(123) &= -\Gamma(231), \\
\nonumber\Theta^3(123) &= -\Gamma(312), \\
\nonumber\Theta^2(1234)&=
-\Gamma(2+3,4,1)\,\Theta^2(1+4,2,3)-\Gamma(2341),\\
\nonumber\Theta^3(1234) &= -\Gamma(3+4,1,2)\,\Theta^2(1+2,3,4)
                -\Gamma(2+3,4,1)\,\Theta^3(1+4,2,3)-\Gamma(3412),\\
\nonumber\Theta^4(1234) &=
-\Gamma(3+4,1,2)\,\Theta^3(1+2,3,4)-\Gamma(4123).\\
\end{align}
When written in terms of their independent momenta they reduce to the simple expressions
\begin{align}
\label{eq:Theta}
\nonumber\Theta^2(123) &= -\Gamma(231), & \Theta^3(123) &= -\Gamma(312), \\
\nonumber\Theta^2(1234)&=
-\frac{\hat{2}}{\hat{1}}\Gamma(1234),& \Theta^3(1234) &=-\frac{\hat{3}}{\hat{1}}\Gamma(1234),\\
\nonumber\Theta^4(1234) &=
-\frac{\hat{4}}{\hat{1}}\Gamma(1234).\\
\end{align}
\section{Symmetries of Scalar Field Theories}
We shall briefly illustrate the extended symmetries of free theories with the example
of complex scalar fields $\varphi$ and $\widetilde{\varphi}$ with action
\begin{equation}\begin{split}
\label{eq:free_action}
S&=\int d^d x \sqrt{-g}\left(g^{\mu\nu}\partial_{\mu}\widetilde{\varphi}\partial_{\nu}\varphi+m^2\widetilde{\varphi}\varphi\right)\\
&=\int d^d x\sqrt{-g}\left(g^{\mu\nu}\widetilde{\varphi}\partial_{\mu}\partial_{\nu}\varphi+m^2\widetilde{\varphi}\varphi\right)\\
&=\int d^d x \sqrt{-g}\widetilde{\varphi}\Omega\varphi
\end{split}\end{equation}
where $\Omega$ is an operator given by $\Omega=\eta^{\mu\nu}\partial_\mu \partial_\nu +m^2$. Consider the transformation $\varphi\rightarrow\varphi+\delta\varphi$ and $\widetilde{\varphi}\rightarrow\widetilde\varphi+\delta\widetilde\varphi$ with $\delta\varphi$ and $\delta\widetilde\varphi $ given by
\begin{equation}
\label{eq:5}
 \delta \varphi(x)=\epsilon \varphi(x_{G})\,,\quad
	\delta\widetilde{\varphi}(x)=-\epsilon \widetilde{\varphi}(x_{G^{-1}})
\end{equation}
resulting from the finite isometry $x\rightarrow x_G$. Now the change in the action is
\begin{equation}
\label{eq:10}
	\delta S=\epsilon \int d^d x \sqrt{-g(x)}\widetilde{\varphi}(x)\Omega(x)\varphi(x_G)-\epsilon \int d^d x \sqrt{-g(x)}\widetilde{\varphi}(x_{G^{-1}})\Omega(x)\varphi(x).
\end{equation}
We are free to apply the isometry $x\rightarrow x_{G}$ to the second integral. By writing $y=x_{G^{-1}}$ and realising that the following is true
\begin{equation}
\nonumber	\sqrt{-g(x)}d^d x=\sqrt{-\grave{g}(y)}d^d y
\end{equation}
then (\ref{eq:10}) becomes
\begin{equation}
\nonumber	\delta S=\epsilon \int d^d x \sqrt{-g(x)}\widetilde{\varphi}(x)\Omega(x)\varphi(x_G)-\epsilon \int d^d y \sqrt{-\grave{g}(y)}\widetilde{\varphi}(y)\Omega(y_{G})\varphi(y_G).
\end{equation}
Since $\Omega$ is an index-less scalar operator we have $\Omega(y_G)=\grave{\Omega}(y)$ and we get
\begin{equation}
\nonumber	\delta S=\epsilon \int d^d x \sqrt{-g(x)}\widetilde{\varphi}(x)\Omega(x)\varphi(x_G)-\epsilon \int d^d y \sqrt{-\grave{g}(y)}\widetilde{\varphi}(y)\grave{\Omega}(y)\varphi(y_G),
\end{equation}
and since $x\rightarrow x_G$ is an isometry we have $\grave{g}=g$ and $\grave{\Omega}=\Omega$ hence we arrive at the conclusion that $\delta S =0$. 

In flat 3+1 spacetime, the isometries are elements of the Poincar\'e group, that is the 6 elements of the Lorentz group and the 4 displacements. For the former 
\begin{equation}
\nonumber	x^{\mu}\rightarrow (x^{\mu})_{G}=\Lambda^{\mu}_{\ \nu}x^{\nu}.
\end{equation}
In the infinitesimal case where $\Lambda^{\mu}_{\ \nu}$ is close to the identity matrix one can write the field transformations as
\begin{eqnarray}
\nonumber	\nonumber\varphi\left(x^{\mu}\right)&\rightarrow&\varphi\left(\delta^{\mu}_{\ \nu}x^{\nu}+\epsilon a^{\mu}_{\ \nu}x^{\nu}\right)\\
\nonumber	\widetilde{\varphi}\left(x^{\mu}\right)&\rightarrow&\widetilde{\varphi}\left(\delta^{\mu}_{\ \nu}x^{\nu}-\epsilon a^{\mu}_{\ \nu}x^{\nu}\right)
\end{eqnarray}
where $a^{\mu}_{\ \nu}$ are the components of an anti-symmetric matrix. We may consider building a finite isometry out of repeated infinitesimal isometries generated by infinitesimal Killing vectors of the spacetime $X(\sigma_x(t))$ where $\sigma_x (t)$ are flows generated by the isometry and $t$ is a parameter \cite{nakahara}. In an infinitesimal case
\begin{eqnarray}
	\nonumber \varphi(x_G)&=&\left(1+\epsilon X^{\mu}\partial_{\mu}+\cdots\right)\varphi(x)\\
\nonumber	&=&\varphi(x)+\epsilon L(x)\varphi(x)+\cdots
\end{eqnarray}
where $L\varphi$ is given by $X^{\mu}(x)\partial_{\mu}\varphi(x)$ and is the Lie derivative of $\varphi$ in the direction of the Killing vector $X(x)$, see \cite{nakahara}. Repeated application of such infinitesimal isometries gives
\begin{eqnarray} \nonumber\varphi(x_G)&=&lim_{N\rightarrow\infty}\left(1+\frac{L(\sigma_x(t_N))}{N}\right)\left(1+\frac{L(\sigma_x(t_{N-1}))}{N}\right)\times\cdots\\
\nonumber&\ &\ \ \ \ \ \ \ \ \ \ \ \ \ \ \ \ \ \ \ \ \ \ \ \ \ \ \ \ \ \ \ \ \ \ \ \ \cdots\times\left(1+\frac{L(\sigma_x(t_0))}{N}\right)\varphi(x)\\
\nonumber &=&Texp\left\{\int_0^t {d\grave{t}L(\sigma_x(t))}\right\}\varphi(x)
\end{eqnarray}
where $T$ is the time ordering operator. We know that 
\begin{equation}
\nonumber	\grave{\varphi}(x)=\varphi(x)+\epsilon\varphi(x_G)=\varphi(x)+\epsilon Texp\left\{\int_0^t {d\grave{t}L(\sigma_x(\grave{t}))}\right\}\varphi(x)
\end{equation}
is a symmetry, where $x(\sigma_x(0))=x$ and $x(\sigma_x (t))=x_G$. Since all terms in a Taylor expansion of the time ordered exponential are linearly independent, each term must itself be a symmetry, so the action must be invariant under the infinitesimal change in the field

\begin{equation}
\nonumber	 \delta_n\varphi(x)=\epsilon\int_0^t{dt_1\int_0^{t_1}{dt_2\dots\int_0^{t_{n-1}}{dt_{n-1}L(t_1)\cdots L(t_n)}}}\varphi(x)
\end{equation}
for $n=0\cdots\infty$. The Lagrangian density changes by a divergence, 
\begin{equation}
\nonumber	\delta^n \mathcal{L}=\epsilon\partial_{\mu}K^{\mu}
\end{equation}
and Noether's theorem gives a corresponding conserved current. For Lorentz transformations $L$ is  $X^{\mu}(x)\partial_{\mu}=a^{\mu}_{\ \nu}x^{\nu}\partial_{\mu}$.
It can be shown that the vector field $K^{\mu}$ is given by
\begin{equation}
\begin{split}
\nonumber	 K^{\mu}_{(n)}=&\int_{0}^{t} dt_1\cdots\int_{0}^{t_{n-1}} dt_n \Big(\eta^{\lambda_{1}\lambda_{n+1}}a^{\mu}_{\ \rho}x^{\rho}(t_1)\partial_{\lambda_{1}}\widetilde{\varphi}\partial_{\lambda_{n+1}}L_{\lambda_2}(t_2)\cdots L_{\lambda_n}(t_n)\varphi\\
	 &-\eta^{\lambda_{1}\lambda_{n+1}}a^{\mu}_{\ \rho}x^{\rho}(t_1)\partial_{\lambda_{1}}L_{\lambda_2}(t_2)\widetilde{\varphi}\partial_{\lambda_{n+1}}L_{\lambda_3}(t_3)\cdots L_{\lambda_n}(t_n)\varphi+\cdots\\
	\ \\
	\ \\ 
	\cdots&+(-1)^{n}\eta^{\lambda_{1}\lambda_{n+1}}a^{\mu}_{\ \rho}x^{\rho}(t_1)\partial_{\lambda_{1}}L_{\lambda_2}(t_2)\cdots L_{\lambda_{n-1}}(t_{n-1})\widetilde{\varphi}\partial_{\lambda_{n+1}}L_{\lambda_n}(t_n)\varphi+\\
	 &+(-1)^{n+1}\eta^{\lambda_{1}\lambda_{n+1}}a^{\mu}_{\ \rho}x^{\rho}(t_1)\partial_{\lambda_{1}}L_{\lambda_2}(t_2)\cdots\ L_{\lambda_n}(t_n)\widetilde{\varphi}\partial_{\lambda_{n+1}}\varphi\Big).
	\end{split}\end{equation}
Notice the abuse of notation for the purpose of abbreviation here. The space-time point $x(t)$ is given by the flow $\sigma$ and should read $\sigma_x(t)$ and $L(t)$ is an abbreviation for $L(X(\sigma_x (t))$ where $X$ is the Killing vector field on the space-time. We could trivially calculate the Noether currents $J^{\mu}_{(n)}$ and prove they are conserved and then the full expression for the current generated by the transformations (\ref{eq:5}) is given by
\begin{equation}
J^{\mu}=\sum_{n}\frac{J^{\mu}_{(n)}}{n!}
\end{equation}
after reintroducing the factors $1/n!$ arising from the Taylor series of the exponential function. This expression is clearly also conserved. 

It is not difficult to generalise this argument to other space time objects by writing the transformation as
\begin{align}
	\nonumber \delta \Phi(x)&=\epsilon U(x)\Phi(x_{G})\\
	\nonumber \delta\widetilde{\Phi}(x)&=-\epsilon \widetilde{\Phi}(x_{G^{-1}})U(x)
\end{align}
where $U(x)$ is some unitary matrix (or possibly a bigger object if $\Phi$ is a type $\left(p,q\right)$ tensor) and in the most general case, the operator $L$ is some generalisation to the normal Lie derivative of the field.

\subsubsection{Lie Algebra}

Let $\varphi(x)$ be a free scalar field and $x\rightarrow x_{G_i}$ be a member of the isometry group $G$
\begin{equation}
\nonumber x\rightarrow x_{G_i}=\Lambda_i x+a
\end{equation}
where $\Lambda$ is the matrix generator of Lorentz boosts and rotations and $a$ is a displacement vector. Then as we have seen a change in the free field $\delta\varphi(x)$ given by 
\begin{equation}
\label{eq:dphi}
\delta\varphi(x)=\epsilon_i\varphi(x_{G_i})
\end{equation}
is a symmetry of the action, eqn (\ref{eq:free_action}). More generally however, it is obvious that linear combinations of eqn (\ref{eq:dphi}), 
\begin{equation}
\nonumber\delta\varphi(x)=\sum_i \epsilon_i\varphi(x_{G_i})
\end{equation}
are also symmetries of eqn (\ref{eq:free_action}) with $\epsilon \in \mathbb{C}$ and the sum being over a discrete subgroup of G for simplicity rather than an integral over the full continuous group. These objects $\delta\varphi(x)$ clearly satisfy the elementary vector space axioms. Two consecutive transformations $\delta_1$ and $\delta_2$ are given by
\begin{equation}
\nonumber\delta_1\delta_2\varphi(x)=\sum_i\sum_j\epsilon_i^1\epsilon_j^2\varphi(x_{G_i G_j})=\sum_i\sum_j\epsilon_i\epsilon_j\varphi(C_{ij}^{\ \ k}x_{G_k})
\end{equation}
with a sum over the index $k$ and with $C_{ij}^{\ \ k}=1$ for one combination of $i$,$j$ and $k$ and zero otherwise. In that sense the $C_{ij}^{\ \ k}$s are a group mulitplication table (or Cayley table) for the discrete subgroup with
\begin{equation}
\label{eq:cayley}
G_iG_j=C_{ij}^{\ \ k}G_k.
\end{equation}
and so $C_{ij}^{\ \ k}$ has only one non vanishing term in the implied sum over $k$. Moreover,  $C_{ij}^{\ \ k}$ can also be taken outside thus,
\begin{equation}
\nonumber\delta_1\delta_2\varphi(x)=\sum_i\sum_j\epsilon_i^1\epsilon_j^2\varphi(x_{G_i G_j})=\sum_i\sum_j\epsilon_i^1\epsilon_j^2 C_{ij}^{\ \ k}\varphi(x_{G_k})
\end{equation}
and the commutator is given by
\begin{equation}
\label{eq:structureconstants}
\left[\delta_1,\delta_2\right]\varphi(x)=\sum_i\sum_j\epsilon_i^1\epsilon_j^2\left(C_{ij}^{\ \ k}-C_{ji}^{\ \ k}\right)\varphi(x_{G_k})=\sum_i\sum_j\epsilon_i^1\epsilon_j^2 f_{ij}^{\ \ k}\delta_k\varphi(x),
\end{equation}
hence satisfying a closure relation. With the transformations $\delta\varphi(x)$ defined this way, the commutators also satisfy the Jacobi identity as follows. Writing out the commutators, we arrive at
\begin{equation}\begin{split}
\nonumber\left[\left[\delta_1,\delta_2\right],\delta_3\right]+\left[\left[\delta_2,\delta_3\right],\delta_1\right]+\left[\left[\delta_3,\delta_1\right],\delta_2\right]\varphi(x)=\\ \sum_i\sum_j\sum_k\epsilon_i^1\epsilon_j^2\epsilon_k^3 \left(f_{ij}^{\ \ l}f_{lk}^{\ \ m}+f_{jk}^{\ \ l}f_{li}^{\ \ m}+f_{ki}^{\ \ l}f_{lj}^{\ \ m}\right)\varphi(x_{G_m})
\end{split}\end{equation}
then writing in terms of $C_{ij}^{\ \ k}$
\begin{equation}\begin{split}
\nonumber \left[\left[\delta_1,\delta_2\right],\delta_3\right]&+\left[\left[\delta_2,\delta_3\right],\delta_1\right]+\left[\left[\delta_3,\delta_1\right],\delta_2\right]\varphi(x)=\\ 
&=\sum_i\sum_j\sum_k\epsilon_i^1\epsilon_j^2\epsilon_k^3\bigg(C_{ij}^{\ \ l} C_{lk}^{\ \ m}-C_{ij}^{\ \ l} C_{kl}^{\ \ m}-C_{ji}^{\ \ l} C_{lk}^{\ \ m}+C_{ji}^{\ \ l} C_{kl}^{\ \ m}\\
&\ \ \ \ \ \ +C_{jk}^{\ \ l} C_{li}^{\ \ m}-C_{jk}^{\ \ l} C_{il}^{\ \ m}-C_{kj}^{\ \ l} C_{li}^{\ \ m}+C_{kj}^{\ \ l} C_{il}^{\ \ m}\\
&\ \ \ \ \ \ +C_{ki}^{\ \ l} C_{lj}^{\ \ m}-C_{ki}^{\ \ l} C_{jl}^{\ \ m}-C_{ik}^{\ \ l} C_{lj}^{\ \ m}+C_{ik}^{\ \ l} C_{jl}^{\ \ m}\bigg)\varphi(x_{G_m})=0,
\end{split}\end{equation}
by using the associativity property $\left(G_iG_j\right)G_k=G_i\left(G_jG_k\right)$ of the group multiplication and eqn (\ref{eq:cayley}). Since the objects $\delta_i$ form a vector space and satisfy commutator closure relations and the Jacobi identity, they are a Lie algebra $g$ over the field $\mathbb{C}$. Since G has has an infinite number of elements the Lie algebra $g$ not only has an infinite number of generators, they are also uncountable due to $G$ being a continous group. However, there exists an infinite number of discrete subgroups of $G$ such as the dihedral subgroups of $SO(3)$ which can be used to form discrete infinite dimensional Lie algebras using the above argument. Algebras constructed in this way are called `group algebras'. (see \cite{james_and_liebeck} for a full discussion on this subject)

In the case when the group $G$ is a discrete subgroup, clearly the dimension of the algebra $g$ is given by
\begin{equation}
\nonumber Dim(g)=\left|G\right|
\end{equation}
where $\left|G\right|$ is the order of the group. In the obvious choice of basis, we have $n=\left|G\right|$ generators of $g$,
\begin{equation}
\label{eq:obvbasis}
\delta_1\varphi(x),\delta_2\varphi(x),\cdots,\delta_n\varphi(x).
\end {equation}
A Lie algebra can be decomposed into the direct sum of a non-Abelian algebra $\bar{g}$ and possibly a trivial Abelian algebra $\mathcal{C}\left(g\right)$ (See \cite{gilmoore}, page 135 and also \cite{fuchs}) refered to as the centre, as follows
\begin{equation}
\label{eq:decomp}
g=\bar{g}\oplus\mathcal{C}\left(g\right).
\end{equation}
The group elements G are distributed amongst conjugacy classes which are subsets of $G$ with mutually orthogonal elements \cite{james_and_liebeck}. 

It turns out that the dimension of $\mathcal{C}\left(g\right)$ equals the number of conjugacy classes of the group $G$ and this is a well known theorem in the subject of group algebras. We shall give a proof that pertains to our application. (For an alternative proof in the more general setting of group algebras, see \cite{james_and_liebeck}.) Let us consider a conjugacy class $C_1$ of $G$ containing $r$ elements say,
\begin{equation}
\nonumber	C_1=\left\{a_1,a_2\cdots a_r\right\}
\end{equation}
and any other element of the group G, say $h$. Now, 
\begin{equation}
\nonumber \delta_{a_i}\delta_h\varphi(x)=\epsilon_{a_i}\epsilon_h\varphi(x_{a_i h})
\end{equation}
and by using the defining relationship between mutually orthogonal elements in a conjugacy class that $hah^{-1}=b$ this equals
\begin{equation}
\nonumber\delta_{a_i}\delta_h\varphi(x)=\epsilon_a\epsilon_h\varphi(x_{h b_i})
\end{equation}
where $b_i$ is also an element (possibly identical to $a_i$) of $C_1$. Also, for any $a_i$ and $a_j$ with $a_i\neq a_j$ we have $b_i\neq b_j$ by the mutually orthogonal property of the conjugacy class. It is possible to construct a generator $\bar{\delta}$ by summing over the elements in $C_1$, 
\begin{equation}
\label{eq:abgen}
\bar{\delta}\varphi(x)=\epsilon_{a}\varphi(x_{a_1})+\epsilon_{a}\varphi(x_{a_2})+\cdots \epsilon_{a}\varphi(x_{a_r})
\end{equation}
so
\begin{equation}
\nonumber\bar{\delta}\delta_h\varphi(x)=\epsilon_{a}\epsilon_h\sum_{r}\varphi\left(x_{a_r h}\right)=\epsilon_{a}\epsilon_{h}\sum_{r}\varphi(x_{h b_r}).
\end{equation}
Now since r runs over all elements in the conjugacy class $C_1$, the right hand sum can be written as
\begin{equation}
\nonumber\epsilon_{a}\epsilon_{h}\sum_{r}\varphi(x_{h b_r})=\epsilon_{a}\epsilon_{h}\sum_{s}\varphi(x_{h a_s})=\delta_h\bar{\delta}\varphi(x)
\end{equation}
giving 
\begin{equation}
\nonumber\left[\bar{\delta},\delta_h\right]=0.
\end{equation}
Now if there are $m$ conjugacy classes $C_1,\cdots,C_m$, this implies that the number of Abelian generators is bigger than or equal to $m$. Equality is proved by assuming we have found $m$ linearly independent Abelian generators given by (\ref{eq:abgen}), $\bar{\delta}_q$ and then constructing an $\left(m+1\right)$th Abelian generator, $\bar{\delta}_{m+1}$, as follows.
\begin{equation}
\nonumber\bar{\delta}_{m+1}\varphi(x)=\epsilon_a\sum_{a_r\in C_1} \lambda_{r1}\varphi(x_{a_{r1}})+\epsilon_a\sum_{a_r\in C_2}\lambda_{r2}\varphi(x_{a_{r2}})+\cdots
\end{equation}
where the sum over $r$ is the sum over the $r$ elements $a_{rq}$ cointained within the conjugacy class $C_q$ and $\lambda_{rq}$ is the coefficient of the the $r$th generator in the $q$th conjugacy class. Then if we take the commutator $\left[\bar{\delta}_{m+1},\delta_h\right]$, it must be zero for for all $\delta_h$ so take the $q$th term in the above sum of $\delta_{m+1}\delta_h$
\begin{equation}
\epsilon_a \epsilon_h\sum_{a_r\in C_q}\lambda_{rq}\varphi(x_{a_{rq}h})=\epsilon_a\epsilon_h\sum_{r}\lambda_{rq}\varphi(x_{hb_{rq}})
\end{equation}
by again using the expression $hah^{-1}=b$. The $a_{rq}$ are all distinct elements so it follows that the $b_{rq}$ are also distinct by the mutual orthogonality property of elements in the conjugacy class. Now relabel the elements $b_{rq}$ as follows, which we can do because we are summing over all elements $a$ (or alternatively $b$) in $C_q$
\begin{equation}
\delta_{m+1}\delta_h\varphi(x)=\cdots+\epsilon_a\epsilon_h\sum_{s}\grave{\lambda}_{sq}\varphi(x_{ha_{sq}})+\cdots
\end{equation}
with $\lambda_{rq}=\grave{\lambda}_{sq}$ and we require, 
\begin{equation}
\delta_{m+1}\delta_h\varphi(x)=\cdots+\epsilon_a\epsilon_h\sum_{s}\grave{\lambda}_{sq}\varphi(x_{ha_{sq}})+\cdots=\cdots+\epsilon_a\epsilon_h\sum_{r}\lambda_{rq}\varphi(x_{ha_{rq}})+\cdots=\delta_h\delta_{m+1}
\end{equation}
which is satisfied only if $\grave{\lambda}_{sq}=\lambda_{rq}$ because the $a_{rq}$ are all distinct linearly independent elements. So we have
\begin{equation}
\nonumber\lambda_{1q}=\lambda_{2q}=\cdots=\lambda_{rq}
\end{equation}
for all conjugacy classes $C_q$, hence $\bar{\delta}_{m+1}$ is in fact a linear combination of $\bar{\delta}_1,\cdots,\bar{\delta}_m$. Hence, if the $n$ elements of $G$ are distributed amongst $m$ conjugacy classes there are exactly $m$ Abelian generators of $\mathcal{C}\left(g\right)$ and the dimension of $\bar{g}$ from eqn (\ref{eq:decomp}) is $n-m$.
\section{Transformation of $A$ and $\overline{A}$}

We shall calculate expressions that leave the the Chalmers-Siegel action $L^{+-}[A]+L^{++-}[A]$ invariant under the transformation $A\rightarrow\grave{A}=A+\delta A$. The operators appearing in the denominators  of $L^{+-}$, $L^{++-}$ and $L^{+-\cdots -}$ are most simply expressed in momentum space. After performing a Fourier transformation on (\ref{eq:field transformations}) we have the following expression absorbing the interaction term on the left hand side into the kinetic term on the right.

\begin{eqnarray}
\label{eq:momentum space action}
 \nonumber tr\int_1 \left\{\bar{p}_1\tilde{p}_1-\hat{p}_1\check{p}_1\right\}\overline{A}_{\bar{1}} A_1-i tr\int_{123} \hat{p}_1\left(\zeta_3-\zeta_2\right)\overline{A}_{\bar{1}}A_{\bar{2}}A_{\bar{3}}(2\pi)^4\delta(p_1+p_2+p_3)\\
=tr\int_1 \left\{\bar{p}_1\tilde{p}_1-\hat{p}_1\check{p}_1\right\}\overline{B}_{\bar{1}}B_1
\end{eqnarray}
where $\zeta_p=\bar{p}/\hat{p}$. In configuration space the isometry is $x\rightarrow x_G=\Lambda x$. Now, Lorentz transformations commute with the Fourier transform, i.e under the isometry $x\rightarrow x_G=\Lambda x$ we have $B^G(p)=B(p_G)=B(\Lambda p)$.  We write the change in the $B$ fields as follows

\begin{eqnarray}
\nonumber\delta B(p)&=&\epsilon B(p_G)\\
\nonumber\delta \overline{B}(p)&=&-\epsilon \overline{B}(p_{G^{-1}}).
\end{eqnarray}

We shall consider the finite isometries in momentum space primarily, however it is instructive to consider the infinitesimal transformations that preserve the quantity $\bar{p}\tilde{p}-\hat{p}\check{p}$ and the finite case will follow.  We have

\begin{equation}
\nonumber\left(\check{p}',\hat{p}',\tilde{p}',\bar{p}'\right)=\left(\check{p},\hat{p},\tilde{p},\bar{p}\right)+\epsilon v_i
\end{equation}
where $v_i$ is given by one of 
\begin{align}
\nonumber &v_1=\left(\bar{p},0,\hat{p},0\right),\  v_2=\left(0,0,-\tilde{p},\bar{p}\right),\  v_3=\left(\tilde{p},0,0,\check{p}\right),\\ \nonumber &v_4=\left(0,\tilde{p},0,\check{p}\right),\  v_5=\left(0,\bar{p},\check{p},0\right),\  v_6=\left(-\check{p},-\hat{p},0,0\right). 
\end{align}
It is simple to substitute these into $\bar{p}'\tilde{p}'-\hat{p}'\check{p}'$ and retrieve $\bar{p}\tilde{p}-\hat{p}\check{p}$ hence showing they have the desired isometry property. By writing the infinitesimal isometries in configuration space, and then Fourier transforming them, we discover that isometries which preserve the quantisation surface $\grave{x}^0=x^0$ also preserve $\hat{p}$ in momentum space. Hence the first three isometries above preserve the constant $x^0$ surfaces. It is also convenient to notice that $\left(\check{p}',\hat{p}',\tilde{p}',\bar{p}'\right)=\left(\check{p},\hat{p},\tilde{p},\bar{p}\right)+\epsilon v_3$ only alters $\check{p}$ and $\bar{p}$, and leaves $\hat{p}$ and $\tilde{p}$ unchanged. Since the coefficients, $\Gamma$ and $\Upsilon$ depend only on $\hat{p}$ and $\tilde{p}$ as mentioned earlier, this will simplify the problem for that one paramater subgroup of isometries. The properties of each of these transformations will be preserved in the finite case also and we shall use this to our advantage by considering only the $\Gamma$ and $\Upsilon$ preserving transformation for the moment but generalising to the other five transformations will turn out to be fairly straight forward. 

\subsection{Transformation of A for the isometry that preserves $\Gamma$ and $\Upsilon$}
Begin with the expression for $A$ in terms of $B$, derived in \cite{ettle_and_morris} and stated earlier,
\begin{equation}
\nonumber	A_1=\sum^{\infty}_{n=2}{\int_{2\cdots n}\Upsilon\left(1 \dots n\right)B_{\overline{2}} \cdots B_{\overline{n}}}.
\end{equation}
The expression for $\delta A$ in terms of $B$ is

\begin{eqnarray}
\label{eq:delA in terms of B}
\nonumber	\delta A_1&=&\sum^{\infty}_{n=2}\sum^{n}_{i=2}{\int_{2\cdots n}\Upsilon\left(1 \dots n\right)B_{\overline{2}} \cdots \delta B_{\overline{i}} \cdots B_{\overline{n}}}\\
&=&\epsilon \sum^{\infty}_{n=2}\sum^{n}_{i=2}{\int_{2\cdots n}\Upsilon\left(1 \dots n\right)B_{\overline{2}} \cdots B_{\overline{i}_G} \cdots B_{\overline{n}}}
\end{eqnarray}
where $B_{\overline{i}^G}$ is shorthand for $B(-p_{i^{G}})$. To the first four orders, this is
\begin{equation}\begin{split}
\label{eq:del A in terms of B}
	\delta A_1=&\epsilon B_{1^G}+\epsilon \int_{23}\Upsilon(123)\left\{B_{\bar{2}^G}B_{\bar{3}}+B_{\bar{2}}B_{\bar{3}^G}\right\}\\
&+\epsilon\int_{234}\Upsilon(1234)\left\{B_{\bar{2}^G}B_{\bar{3}}B_{\bar{4}}+B_{\bar{2}}B_{\bar{3}^G}B_{\bar{4}}+B_{\bar{2}}B_{\bar{3}}B_{\bar{4}^G}\right\}\\
&+\epsilon\int_{2345}\Upsilon(12345)\left\{B_{\bar{2}^G}B_{\bar{3}}B_{\bar{4}}B_{\bar{5}}+B_{\bar{2}}B_{\bar{3}^G}B_{\bar{4}}B_{\bar{5}}+B_{\bar{2}}B_{\bar{3}}B_{\bar{4}^G}B_{\bar{5}}+B_{\bar{2}}B_{\bar{3}}B_{\bar{4}}B_{\bar{5}^G}\right\}\\
&+\cdots
\end{split}\end{equation}
Temporarily re-instating the delta functions, we can now substitute the inverse expression $B$ in terms of $A$ given by
\begin{equation}
\nonumber	B_1=\sum^{\infty}_{n=2}{\int_{2\cdots n}\Gamma\left(1 \dots n\right)A_{\overline{2}} \cdots A_{\overline{n}}}(2\pi)^4\delta^4\left(p_1+\cdots+p_n\right).
\end{equation}
There is the added complication that we are evaluating $B(p)$ at $B(p_G)$ but this is dealt with using the property of the delta function that $\delta^4\left(\Lambda p_1+\cdots+\Lambda p_n\right)=\delta^4\left(p_1+\cdots+p_n\right)$. $B_{1^G}$ is given by
\begin{equation}
\nonumber	B_{1^G}=\sum^{\infty}_{n=2}{\int_{2\cdots n}\Gamma\left(1_G, 2, \dots ,n\right)A_{\overline{2}} \cdots A_{\overline{n}}}(2\pi)^4\delta^4\left(p_1^G+\cdots+p_n\right)
\end{equation}
and we can change variables under the integrals using the isometry $p \rightarrow p_G$ to get the following expression. It is an isometry so the Jacobian of the transformation is 1,
\begin{equation}
\nonumber	B_{1^G}=\sum^{\infty}_{n=2}{\int_{2\cdots n}\Gamma\left(1_G, 2_G, \dots ,n_G\right)A_{\overline{2}^G} \cdots A_{\overline{n}^G}}(2\pi)^4\delta^4\left(p_1^G+\cdots+p_n^G\right)
\end{equation} 
which is
\begin{equation}
\nonumber	B_{1^G}=\sum^{\infty}_{n=2}{\int_{2\cdots n}\Gamma\left(1_G, 2_G, \dots ,n_G\right)A_{\overline{2}^G} \cdots A_{\overline{n}^G}}(2\pi)^4\delta^4\left(p_1+\cdots+p_n\right)
\end{equation} 
using the stated property of the delta function. Further, seeing as for the moment we are considering the one transformation that leaves $\Gamma$ and $\Upsilon$ invariant we have
\begin{equation}
\nonumber	B_{1^G}=\sum^{\infty}_{n=2}{\int_{2\cdots n}\Gamma\left(1, 2, \dots ,n\right)A_{\overline{2}^G} \cdots A_{\overline{n}^G}}(2\pi)^4\delta^4\left(p_1+\cdots+p_n\right).
\end{equation} 
Performing the substitution into (\ref{eq:del A in terms of B}) and working up to fourth order only for now, taking care with delta functions, maintaining the order of the A fields and labeling the momentum arguments we get a somewhat nasty looking expression which is included in appendix (\ref{ap:A}). When like terms are collected and their coefficients calculated in terms of independent momenta the expression simplifies into something more tangible. We shall collect terms order by order.  First order is trivial, we get $\delta A=\epsilon A_{1^G}+\cdots$.
The next two orders in A are given below and the more cumbersome fourth order result is included in appendix (\ref{ap:A}) by (\ref{eq:dA4thorder}).

\paragraph{Second Order}
\begin{equation}
\nonumber	\delta A_{1}=\epsilon A_{1^G}+\epsilon i\int_{23}\left\{\frac{\hat{1}\ A_{\bar{2}^G}A_{\bar{3}^G}}{(23)}-\frac{\hat{1}\ A_{\bar{2}^G}A_{\bar{3}}}{(23)}-\frac{\hat{1}\ A_{\bar{2}}A_{\bar{3}^G}}{(23)}\right\}+\cdots
\end{equation}
\paragraph{Third Order}
\begin{equation}\begin{split}
\nonumber \cdots+\epsilon\int_{234}\bigg\{&\frac{\hat{1}\hat{q}\ A_{\bar{2}^G}A_{\bar{3}^G}A_{\bar{4}^G}}{(q,2)(q,2+3)}+\frac{\hat{1}\hat{q}\ A_{\bar{2}^G}A_{\bar{3}^G}A_{\bar{4}}}{(q,2)(q,4)}+\frac{\hat{1}\hat{q}\ A_{\bar{2}}A_{\bar{3}^G}A_{\bar{4}^G}}{(q,3)(q,1)}\\
&+\frac{\hat{1}\hat{q}\ A_{\bar{2}^G}A_{\bar{3}}A_{\bar{4}}}{(q,3)(q,3+4)}+\frac{\hat{1}\hat{q}\ A_{\bar{2}}A_{\bar{3}^G}A_{\bar{4}}}{(q,4)(q,4+1)}+\frac{\hat{1}\hat{q}\ A_{\bar{2}}A_{\bar{3}}A_{\bar{4}^G}}{(q,1)(q,1+2)}\bigg\}+\cdots
\end{split}\end{equation}
where for any term with $A_{\bar{2}}\cdots A_{\bar{i}^G} \cdots A_{\bar{j}^G} \cdots A_{\bar{n}}$, $q$ is defined to be $q=p_i+ \cdots +p_j$.
We may now be tempted to hypothesize the full expression. We write
\begin{eqnarray}
\label{eq:deltaA}
	\nonumber\delta A_1=-\epsilon\sum_{n=2}^{\infty}\sum_{i=2}^{n}\sum_{j=i}^n\int_{2\cdots n}\frac{\hat{1}}{\hat{q}}\Gamma(q,i,\cdots,j)\Gamma(q,j+1,\cdots,n,1\cdots,i-1)\times\\
\nonumber	\times A_{\bar{2}}\cdots A_{\bar{i}^G}\cdots A_{\bar{j}^G}\cdots A_{\bar{n}}\\
\end{eqnarray}
where $\Gamma$ is given by (\ref{eq:Gamma}) and $q=p_i+\cdots +p_j$ as before. Notice this is a cyclic insertion of the momentum arguments into the product of the $\Gamma$s. It is a simple matter to check that this expression does indeed generate the first, second, third and fourth order terms. A diagrammatic representation of this expression is extremely beneficial where we attach $A$ fields to the external legs of a momentum flow diagram whose momenta flow out of the two vertices $\Gamma$ connected by an internal line with momentum $q$ and summing over all diagrams, fig (1),
\begin{figure}[h]
\begin{center}
    \includegraphics{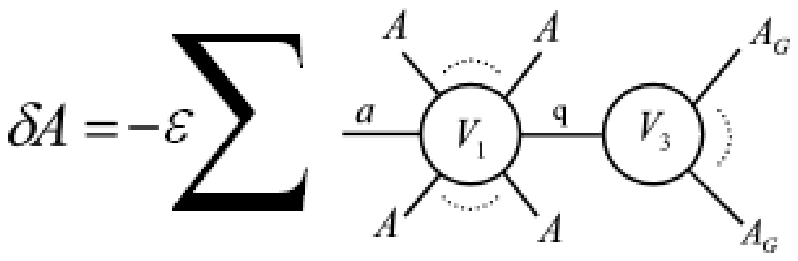}
    \caption{$\delta A$}
 \end{center}
\end{figure}
where the vertices labeled $V_1$ and $V_3$ are expressed in terms of $k$, $q$ and $\Gamma$ which we are forcing to be invariant at the moment and are given explicitly by
\begin{equation}
\nonumber V_1=\frac{\hat{a}}{\hat{q}}\Gamma\ \ \ \ \ \ \ V_3=\Gamma.
\end{equation}

It is reasonable to expect the transformations of the field $A$ satisfy the same algebra and in fact this is easy to prove. From eqn (\ref{eq:delA in terms of B}) we have
\begin{equation}
\nonumber\delta_2 A_1=\sum_{q=2}^n\sum_{n=2}^\infty\int_{2\cdots n} \Upsilon(1\cdots n)B_{\bar{2}}\cdots\delta_2 B_{\bar{q}}\cdots B_{\bar{n}}.
\end{equation}
Two consecutive transformations are given by
\begin{equation}
\nonumber\delta_i\delta_j A_1=\sum_{p=2}^n\sum_{q=2}^n\sum_{n=2}^\infty\int_{2\cdots n} \Upsilon(1\cdots n)B_{\bar{2}}\cdots\delta_i B_{\bar{p}}\cdots\delta_j B_{\bar{q}}\cdots B_{\bar{n}}
\end{equation}
and the commutator is
\begin{equation}\begin{split}
\nonumber\left[\delta_i,\delta_j\right] A_1=\sum_{p=2}^n\sum_{q=2}^n\sum_{n=2}^\infty\int_{2\cdots n}\bigg( \Upsilon(1\cdots n)B_{\bar{2}}\cdots\delta_i B_{\bar{p}}\cdots\delta_j B_{\bar{q}}\cdots B_{\bar{n}}-\\
- \Upsilon(1\cdots n)B_{\bar{2}}\cdots\delta_j B_{\bar{p}}\cdots\delta_i B_{\bar{q}}\cdots B_{\bar{n}}\bigg).
\end{split}\end{equation}
After summing over $p$ and $q$ all terms are zero except those for which $p=q$, leaving only
\begin{equation}\begin{split}
\nonumber\left[\delta_i,\delta_j\right] A_1=\sum_{p=2}^n\sum_{n=2}^\infty\int_{2\cdots n}\Upsilon(1\cdots n)\left(B_{\bar{2}}\cdots\delta_i\delta_j B_{\bar{p}}\cdots B_{\bar{n}}-B_{\bar{2}}\cdots\delta_j\delta_i B_{\bar{p}}\cdots B_{\bar{n}}\right)=\\
=\sum_{p=2}^n\sum_{n=2}^\infty\int_{2\cdots n}\Upsilon(1\cdots n)\left(B_{\bar{2}}\cdots B_{\bar{p}^{G_{ji}}}\cdots B_{\bar{n}}-B_{\bar{2}}\cdots B_{\bar{p}^{G_{ij}}}\cdots B_{\bar{n}}\right)=\\
=\left(C_{ij}^{\ \ k}-C_{ji}^{\ \ k}\right)\sum_{p=2}^n\sum_{n=2}^\infty\int_{2\cdots n}\Upsilon(1\cdots n)B_{\bar{2}}\cdots B_{\bar{p}^{G_k}}\cdots B_{\bar{n}}=\\
=\left(C_{ij}^{\ \ k}-C_{ji}^{\ \ k}\right)\delta_k A_1
\end{split}\end{equation}
which has the same structure constants $f_{ij}^{\ \ k}=\left(C_{ij}^{\ \ k}-C_{ji}^{\ \ k}\right)$ as the commutators in the free theory, eqn (\ref{eq:structureconstants}) thus identifying the algebra unambiguously with that of the free theory. It makes sense therefore to study the algebra of the transformations given by eqn (\ref{eq:deltaA}) in the free theory knowing that the algebra in the less trivial self dual Yang-Mills setting will be the same.

\subsection{Transformation of $\overline{A}$ for the isometry that preserves $\Gamma$ and $\Upsilon$}
The expression for the change in the conjugate field is not dissimilar, although the expansion is significantly more detailed.  The change in the free $B$ field is defined as
\begin{equation}
	\nonumber\delta \overline{B}(p)=-\epsilon \overline{B}(p_{G^{-1}}).
\end{equation}
Let us consider the change in $\overline{A}$ in terms of $B$. 
\begin{equation}\begin{split}
\label{eq:delAbar in terms of B}	\nonumber\delta\overline{A}_1=\delta\overline{B}_1-\int_{23}\bigg\{&\frac{\hat{2}}{\hat{1}}\Xi^2(123)\delta\overline{B}_{\bar{2}}B_{\bar{3}}-\frac{\hat{2}}{\hat{1}}\Xi^2(123)\overline{B}_{\bar{2}}\delta B_{\bar{3}}-\\
&-\frac{\hat{3}}{\hat{1}}\Xi^3(123)\delta B_{\bar{2}}\overline B_{\bar{3}}-\frac{\hat{3}}{\hat{1}}\Xi^3(123)B_{\bar{2}}\overline \delta B_{\bar{3}}\bigg\}-\cdots.
\end{split}\end{equation}
Now substitute the change in the $B$ fields, $\delta B_p=\epsilon B_{p^{G^{-1}}}$ and $\delta\overline{B}=-\epsilon \overline{B}_{p^{G^{-1}}}$

\begin{equation}\begin{split}	\nonumber\delta\overline{A}_1=&-\epsilon\overline{B}_{1^{G^{-1}}}+\\
&+\epsilon\int_{23}\bigg\{\frac{\hat{2}}{\hat{1}}\Xi^2(123)\overline{B}_{\bar{2}^{G^{-1}}}B_{\bar{3}}-\frac{\hat{2}}{\hat{1}}\Xi^2(123)\overline{B}_{\bar{2}}B_{\bar{3}^G}-\\
&\ \ \ \ \ \ \ \ \ \ -\frac{\hat{3}}{\hat{1}}\Xi^3(123) B_{\bar{2}^G}\overline B_{\bar{3}}+\epsilon\int_{23}\frac{\hat{3}}{\hat{1}}\Xi^3(123)B_{\bar{2}}\overline B_{\bar{3}^{G^{-1}}}\bigg\}\\
 &+\epsilon\int_{234}\bigg\{\frac{\hat{2}}{\hat{1}}\Xi^2(1234)\overline{B}_{\bar{2}^{G^{-1}}}B_{\bar{3}}B_{\bar{4}}-\frac{\hat{2}}{\hat{1}}\Xi^2(1234)\overline{B}_{\bar{2}}B_{\bar{3}^G}B_{\bar{4}}-\frac{\hat{2}}{\hat{1}}\Xi^2(1234)\overline{B}_{\bar{2}}B_{\bar{3}}B_{\bar{4}^G}\\
 &\ \ \ \ \ \ \ \ \  -\frac{\hat{3}}{\hat{1}}\Xi^3(1234)B_{\bar{2}^{G}}\overline{B}_{\bar{3}}B_{\bar{4}}+\frac{\hat{3}}{\hat{1}}\Xi^3(1234)_{\bar{2}}\overline{B}_{\bar{3}^{G^{-1}}}B_{\bar{4}}-\frac{\hat{3}}{\hat{1}}\Xi^3(1234)B_{\bar{2}}\overline{B}_{\bar{3}}B_{\bar{4}^G}\\
 &\ \ \ \ \ \ \ \ \  -\frac{\hat{4}}{\hat{1}}\Xi^4(1234)B_{\bar{2}^G}B_{\bar{3}}\overline{B}_{\bar{4}}-\frac{\hat{4}}{\hat{1}}\Xi^4(1234)B_{\bar{2}}B_{\bar{3}^G}\overline{B}_{\bar{4}}+\frac{\hat{4}}{\hat{1}}\Xi^4(1234)B_{\bar{2}}B_{\bar{3}}\overline{B}_{\bar{4}^{G^{-1}}}\bigg\}+\cdots
\end{split}\end{equation}
to third order. In a similar fashion to the previous calculation, we substitute the inverse expressions, $B[A]$ and  $\overline{B}[A,\overline{A}]$ which is given by the expansion in appendix (\ref{ap:B}). Again, we shall collect terms order by order and we shall see that we have already done most of the work when calculating the coefficients earlier. First order is again trivial, we get $\delta \overline{A}_1=-\epsilon\overline{A}_{1^{G^-1}}+\cdots$. At second order we can pick out the terms and express $\Xi$ and $\Theta$ in terms of independent momenta, no extra calculation is required and the result is given below. The third order result is given in appendix (\ref{ap:B}) by eqn (\ref{eq:dbarA3rdorder})
\paragraph{Second Order}
\begin{equation}\begin{split}	\nonumber\delta\overline{A}_1=-\epsilon\overline{A}_{1^{G^{-1}}}-\epsilon\int_{23}i\bigg\{&\frac{\hat{2}}{\hat{1}}\frac{\hat{2}}{(31)}\overline{A}_{\bar{2}^{G^{-1}}}A_{\bar{3}^{G^{-1}}}-\frac{\hat{3}}{\hat{1}}\frac{\hat{3}}{(12)}A_{\bar{2}^{G^{-1}}}\overline{A}_{\bar{3}^{G^{-1}}}\\
&+\frac{\hat{2}}{\hat{1}}\frac{\hat{2}}{(31)}\overline{A}_{\bar{2}^{G^{-1}}}A_{\bar{3}}-\frac{\hat{2}}{\hat{1}}\frac{\hat{2}}{(31)}\overline{A}_{\bar{2}}A_{\bar{3}^G}\\
&-\frac{\hat{3}}{\hat{1}}\frac{\hat{3}}{(12)}A_{\bar{2}^G}\overline{A}_{\bar{3}}+\frac{\hat{3}}{\hat{1}}i\frac{\hat{3}}{(12)}A_{\bar{2}}\overline{A}_{\bar{3}^{G^{-1}}}\bigg\}+\cdots.
\end{split}\end{equation}
We hypothesize the full expression is
\begin{eqnarray}	
\label{eq:deltaAbar}
\nonumber \delta \overline{A}_1=-\epsilon\sum_{n=2}^{\infty}\sum_{k=2}^n\sum_{i=2}^{k-1}\sum_{j=i}^{k-1}\int_{2\cdots n}\frac{\hat{k}^2}{\hat{1}\hat{q}}\Gamma(q,i,\cdots,j)\Gamma(q,j+1,\cdots,n,1,\cdots,i-1)\times\\
\nonumber\times A_{\bar{2}}\cdots A_{\bar{i}^G}\cdots A_{\bar{j}^G}\cdots \overline{A}_{\bar{k}}\cdots A_{\bar{n}}\\
\nonumber+\epsilon\sum_{n=2}^{\infty}\sum_{k=2}^n\sum_{i=2}^{k}\sum_{j=k}^{n}\int_{2 \cdots n}\frac{\hat{k}^2}{\hat{1}\hat{q}}\Gamma(q,i,\cdots,j)\Gamma(q,j+1,\cdots,n,1,\cdots,i-1)\times\\
\nonumber\times A_{\bar{2}}\cdots A_{\bar{i}^{G^{-1}}}\cdots \overline{A}_{\bar{k}^{G^{-1}}}\cdots A_{\bar{j}^{G^{-1}}}A_{\bar{n}}\\
\nonumber-\epsilon\sum_{n=2}^{\infty}\sum_{k=2}^n\sum_{i=k+1}^{n}\sum_{j=i}^{n}\int_{2 \cdots n}\frac{\hat{k}^2}{\hat{1}\hat{q}}\Gamma(q,i,\cdots,j)\Gamma(q,j+1,\cdots,n,1,\cdots,i-1)\times\\
\times A_{\bar{2}}\cdots \overline{A}_{\bar{k}}\cdots A_{\bar{i}^G}\cdots A_{\bar{j}^G}\cdots A_{\bar{n}}.
\end{eqnarray}
It is possible to verify that this expression reproduces first, second and third order terms and again, encoding the expression in a diagrammatic fashion is beneficial fig (2). We have a series of similar diagrams to fig (1) but with cyclic permutations of the $\overline{A}$ field over diagrams in the series.
\begin{figure}[h]
 \begin{center}
    \includegraphics{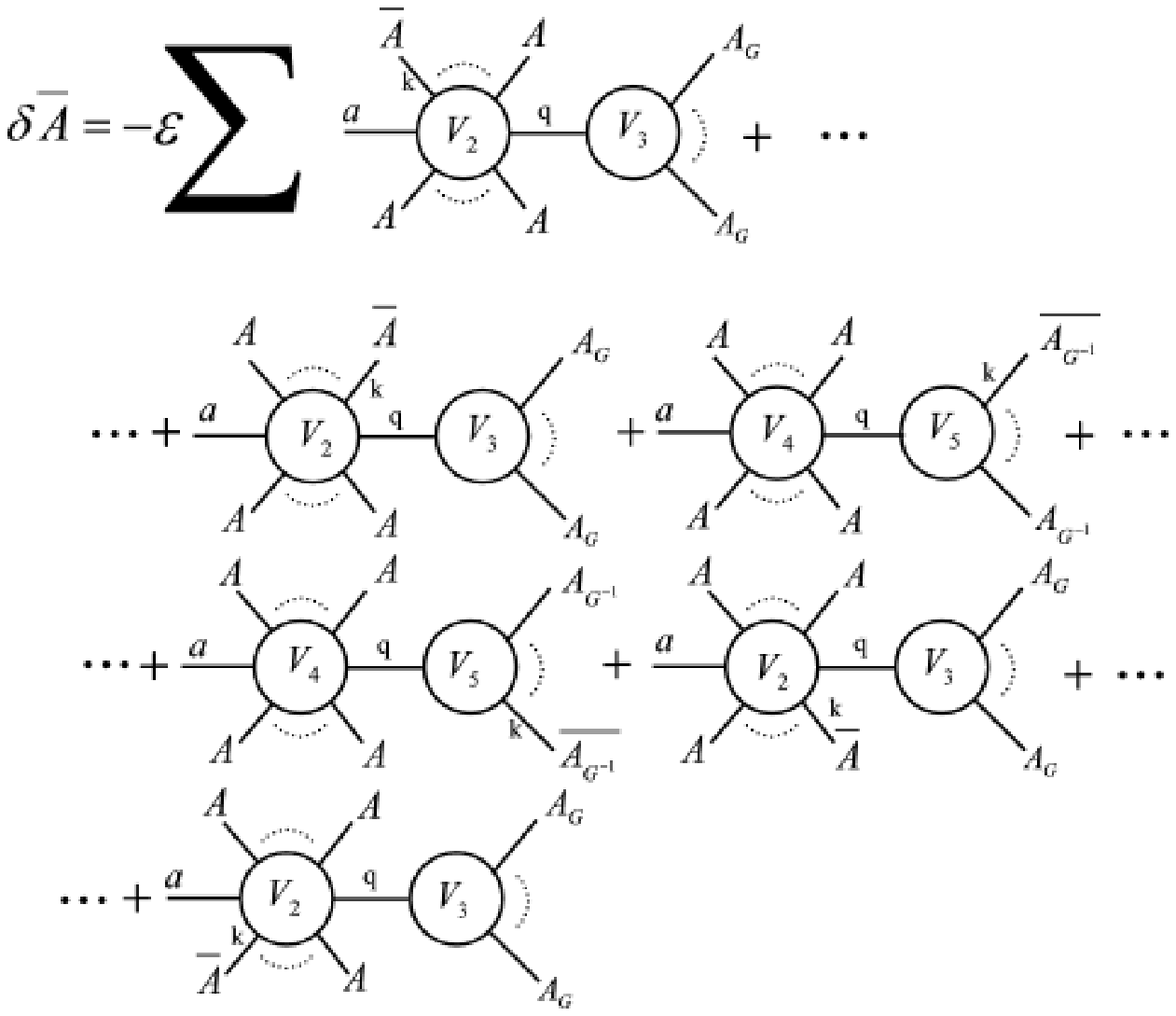}
    \caption{$\delta \overline{A}$}
 \end{center}
\end{figure}
Notice also, the distribution of $A_G$ and $A_{G^{-1}}$ legs in relation to the position of the conjugate field. The transformed legs all flow out of the right hand vertex in each diagram. If the conjugate field is attached to the right hand vertex, then all fields attached to the right hand vertex are transformed as $A_{G^{-1}}$. If the congugate field is not connected to the right hand vertex but rather the left vertex, then the fields attached to it are transformed as $A_G$. The symbol $k$ labels the position of the conjugate field and $a$ labels the position of the `in-coming' leg of the diagram. The vertices, $V_2$, $V_4$ and $V_5$ are given by
\begin{equation}
 \nonumber  V_2=\frac{\hat{a}\hat{k}^2}{\hat{q}\hat{a}^2}\Gamma\ \ \ \ \ \ \ V_4=\frac{\hat{q}}{\hat{a}}\Gamma\ \ \ \ \ \ \ \ V_5=\frac{\hat{k}^2}{\hat{q}^2}\Gamma.\ \ \ \ \ \ \ \
\end{equation}
We now have a conjecture for $\delta A$ and $\delta\overline{A}$ for the transformation which leaves $\check{p}$ and $\tilde{p}$ unchanged. We shall not prove this now but instead we shall hypothesise the most general case by considering the remaining five Lorentz transformations using results thus far and prove that they leave the Chalmers-Siegel action invariant.
\subsection{Most general transformation using the full Lorentz group}
Up to now we have considered the one isometry that leaves the coefficients $\Gamma$ and $\Upsilon$ invariant, namely
\begin{equation}
\nonumber\left(\check{p}',\hat{p}',\tilde{p}',\bar{p}'\right)=\left(\check{p},\hat{p},\tilde{p},\bar{p}\right)+\epsilon\left(\tilde{p},0,0,\check{p}\right).
\end{equation}
Generally of course, $\Gamma$ is not invariant under the six parameter independent Lorentz transformations. In the case of the isometries that preserve the quantisation surface (surfaces of constant $x_0$ or equivalently constant $\hat{p}$), the prefactors $\hat{1}/\hat{q}$ and $\hat{k}^2/\hat{1}\hat{q}$ appearing in (\ref{eq:deltaA}) and (\ref{eq:deltaAbar}) respectively are invariant but more generally these also transform under the full Lorentz group. Writing the vertex factors in the diagrams as we have done in fig (1) and fig (2) it strongly suggests the form of the most general expressions as fig (3), with transformed expressions in the appropriate vertices.
\begin{figure}[h]
\begin{center}
 \includegraphics{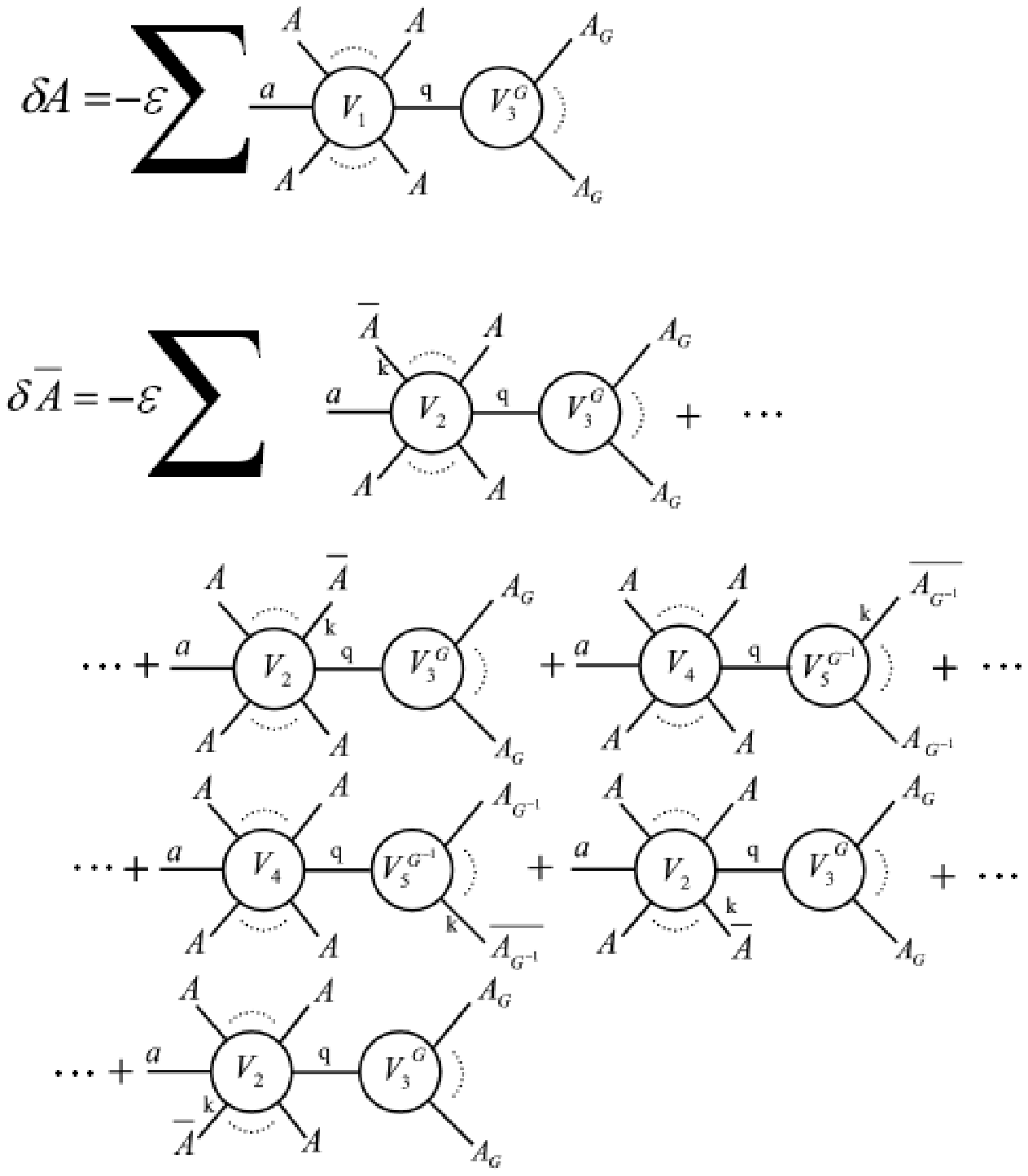}\\
    \caption{Expressions for $\delta A$ and $\delta \overline{A}$ for the full lorentz group}
\end{center}
\end{figure}
The proof of these invariances is obtained by substituting them into the change in action, (\ref{eq:momentum space action}). Algebraically, performing the variation of the action gives us.
\begin{eqnarray}
\label{eq:deltaS}
 \nonumber \delta S= tr\int_1 \left\{\bar{p}_1\tilde{p}_1-\hat{p}_1\check{p}_1\right\}(\delta\overline{A}_{\bar{1}}) A_1+tr\int_1 \left\{\bar{p}_1\tilde{p}_1-\hat{p}_1\check{p}_1\right\}\overline{A}_{\bar{1}} (\delta A_1)\\
 \nonumber-i tr\int_{123} \hat{p}_1\left(\zeta_3-\zeta_2\right)(\delta\overline{A}_{\bar{1}})A_{\bar{2}}A_{\bar{3}}-i tr\int_{123} \hat{p}_1\left(\zeta_3-\zeta_2\right)\overline{A}_{\bar{1}}(\delta A_{\bar{2}})A_{\bar{3}}\\
 -i tr\int_{123} \hat{p}_1\left(\zeta_3-\zeta_2\right)\overline{A}_{\bar{1}}A_{\bar{2}}(\delta A_{\bar{3}}).
\end{eqnarray}
It will be easier to separate out the free and interacting parts of the action and consider their diagrams separately, i.e $\delta S=\delta S_F+\delta S_I$. Each piece reduces to a simpler algebraic expression by considering their diagrammatic expansions, and taken together they will sum to zero. Diagrammatically, the free action $S_F$ is a sum over two, two point vertices and the interacting part is a sum over three point vertices as shown in fig (4).
\begin{figure}[h]
\begin{center}
   \includegraphics{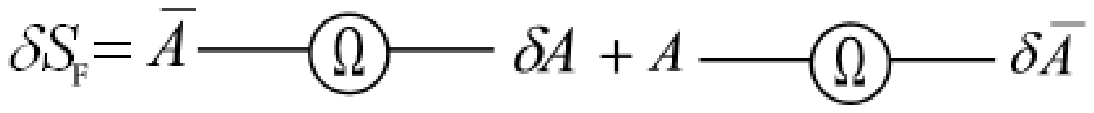}\\
    \includegraphics{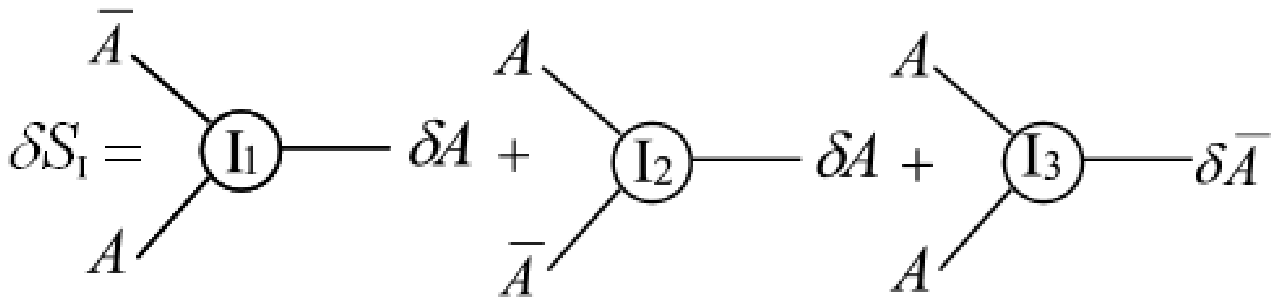}
    \caption{$\delta S$}
\end{center}   
\end{figure}
where $\Omega=\left\{\bar{p}_1\tilde{p}_1-\hat{p}_1\check{p}_1\right\}$. Recall that $\Omega$ is invariant under isometries $x\rightarrow x_G$ whereas the expression $I$ appearing in fig (4) is not invariant. The sum over all diagrams for $\delta S_F$ is relatively straight forward. We have fig (5)
\begin{figure}[h]
\begin{center}
    \includegraphics{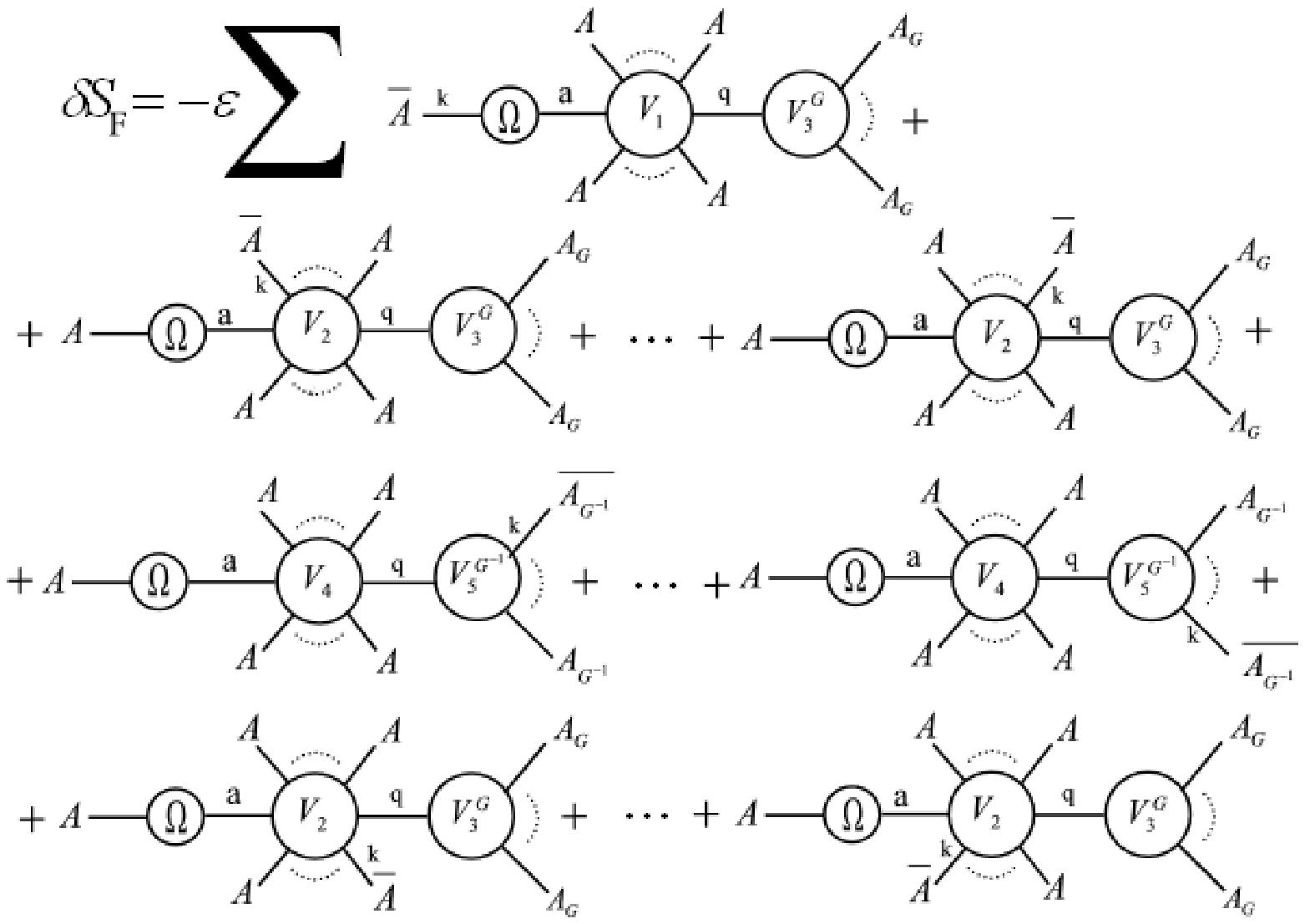}
    \caption{Change in the free part of the action $\delta S_F$}
\end{center}
\end{figure}
where again, the symbol k labels the leg to which the conjugate field is attached and we are free to label the momentum of this leg, $p_1$. Now we can apply the isometry $x\rightarrow x_G$ to diagrams containing $A_{G^{-1}}$, fig (6).
\begin{figure}[h]
\begin{center}
    \includegraphics{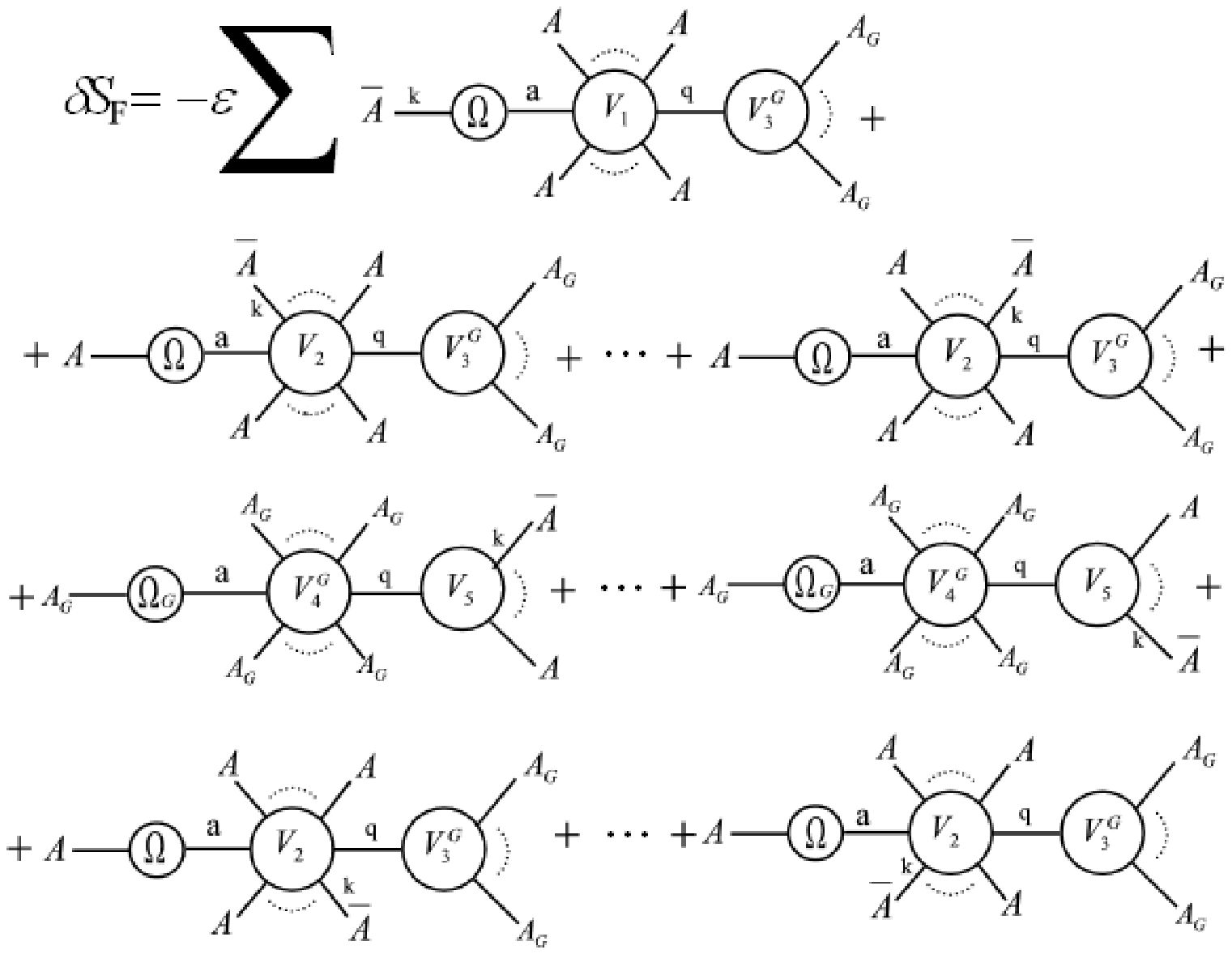}
    \caption{Change in the free part of the action $\delta S_F$}
 \end{center}
\end{figure}
Notice the cyclic permutation of the $\overline{A}$ field which is equivalent to a cyclic permutation of the two point vertex, $\Omega$. Algebraically then, these diagrams reduce to the following expression, involving a product of $\Gamma$s and sum over $k$ of $\Omega(k)$ arising from cyclically permuting the $\Omega$ vertex over the out going legs of the $V$ and $V^G$ vertices,
\begin{eqnarray}
\nonumber	\delta S_F=\epsilon\sum_{n=2}^{\infty}\sum_{i=2}^{n}\sum_{j=i}^{n}\int_{1 \cdots n}X_{i,j}\left(1,\cdots,n\right)\overline{A}_{\bar{1}}\cdots A_{\bar{i}^G}\cdots A_{\bar{j}^G}\cdots A_{\bar{n}}
\end{eqnarray}
where the coefficient $X_{i,j}$ is given by 
\begin{equation}\begin{split} \nonumber &X_{i,j}\left(1,\cdots,n\right)=\\
&\ \ \ \ \ \ =-\frac{\hat{1}^2}{\hat{q}^2}\Big(\frac{\hat{q}}{\hat{1}}\Omega_1+\cdots+\frac{\hat{q}}{\widehat{i-1}}\Omega_{i-1}+\frac{\hat{q}^G}{\hat{i}^G}\Omega_i^G+\cdots+\frac{\hat{q}^G}{\hat{j}^G}\Omega_j^G+\frac{\hat{q}}{\widehat{j+1}}\Omega_{j+1}+\cdots+\frac{\hat{q}}{\hat{n}}\Omega_n\Big)\times\\
&\ \ \ \ \ \ \ \ \ \ \ \ \ \ \ \ \ \ \ \ \ \ \ \ \ \ \ \ \ \ \ \ \ \ \ \ \ \ \ \ \ \ \ \ \ \ \times\Gamma\left(q^G,i^G,\cdots,j^G\right)\Gamma\left(q,j+1,\cdots,i-1\right).
\end{split}\end{equation}
The interacting part is similar except with the cyclic permutation of a three point vertex around the $V$ vertices as opposed to the two point vertex. We attach the diagrams $\delta A$ and $\delta\overline A$ from fig (3) respectively to $\delta S_I$ as shown in fig (7).
\begin{figure}[h]
\begin{center}
     \includegraphics{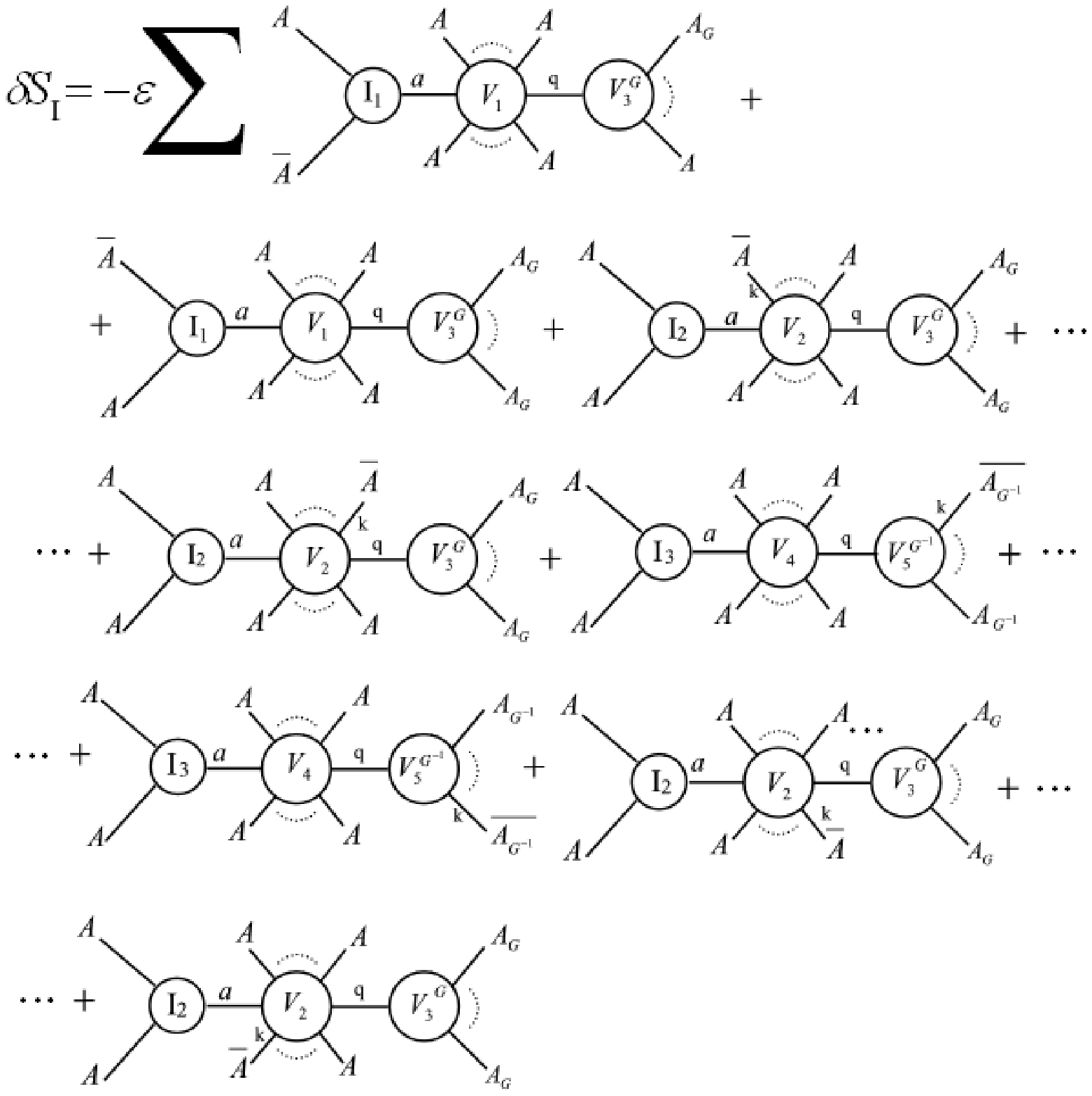}
    \caption{Change in the interacting part of the action $\delta S_I$}
 \end{center}
\end{figure}
where $k$, again labels the conjugate field which we are free to label as momentum $p_1$ and $a$ labels the leg to which the vertex I is attached. We proceed to reverse the isometry from the appropriate diagrams, fig (8).
\begin{figure}[h]
\begin{center}
    \includegraphics{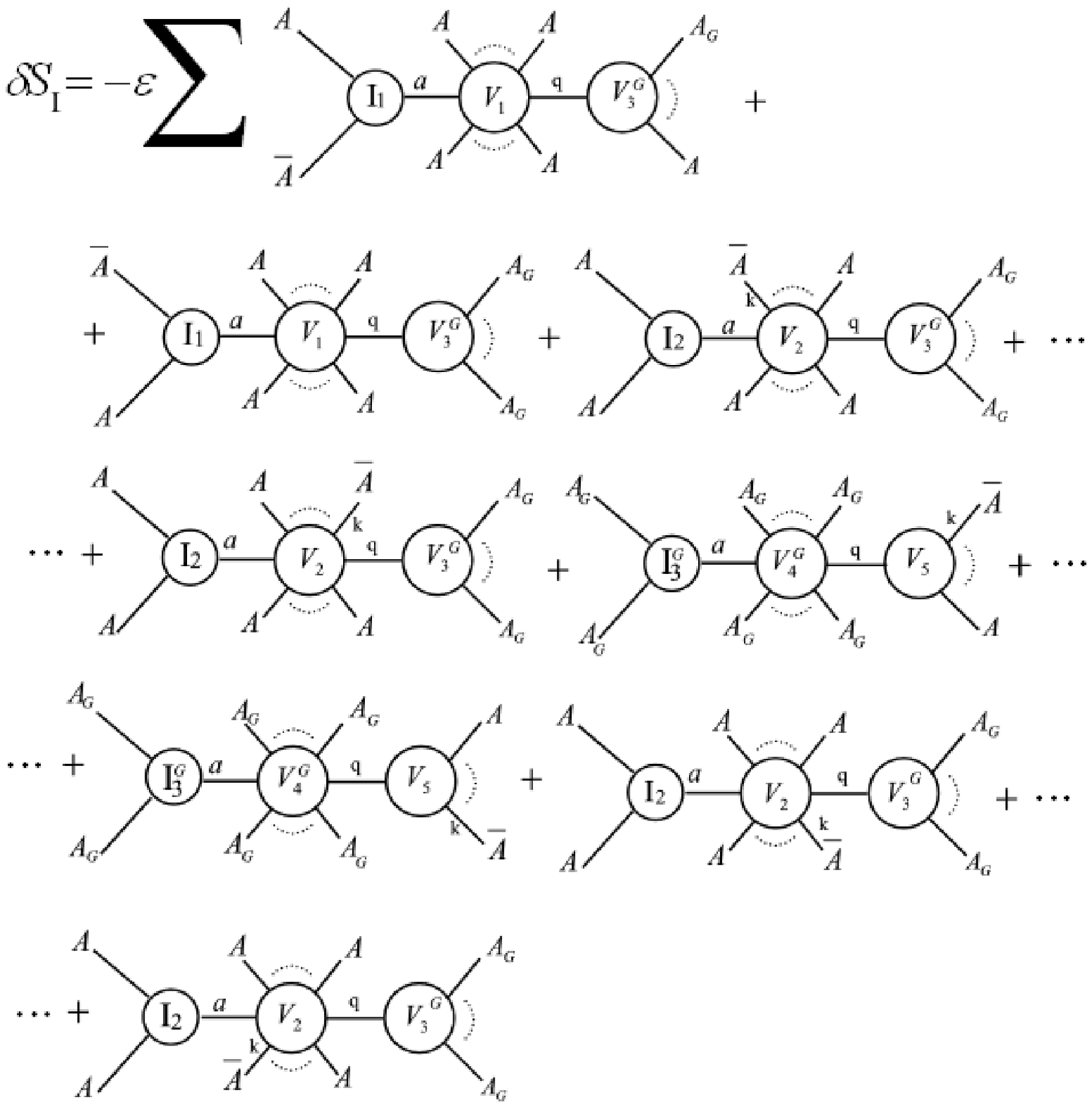}
    \caption{Change in the interacting part of the action $\delta S_I$}
\end{center}
\end{figure}
If the leg $k$ has momentum $p_1$, these diagrams are interpreted as a cyclic permutation of the three point vertex, $I$.
Adding cyclic contributions together gives.
\begin{eqnarray}
	\nonumber	\delta S_I=\epsilon\sum_{n=2}^{\infty}\sum_{i=2}^{n}\sum_{j=i}^{n}\int_{1\cdots n}Y_{i,j}\left(1,\cdots,n\right)\overline{A}_{\bar{1}}\cdots A_{\bar{i}^G}\cdots A_{\bar{j}^G} A_{\bar{n}}
\end{eqnarray}
with the coefficient $Y_{i,j}$ given by
\begin{eqnarray}	\nonumber Y_{i,j}\left(1,\cdots,n\right)=-i\frac{\hat{1}^2}{i\hat{q}^2}\left(\sum_{k=i}^{j-1}\frac{\left\{k^G,(k+1)^G\right\}}{\hat{k}^G\widehat{k+1}^G}\left(q^G,P_{i,k}^G\right)+\sum_{k=j+1}^{i-2}\frac{\left\{k,k+1\right\}}{\hat{k}\widehat{k+1}}\left(q,P_{i,k}\right)\right)\times\\
\nonumber	\times\Gamma\left(q^G,i^G,\cdots,j^G\right)\Gamma\left(q,j+1,\cdots,i-1\right).
\end{eqnarray}
The notation, $P_{i,k}$ means $P_{i,k}=p_i+\cdots+p_k$. Expanding out the summations in the brackets, either explicitly or by comparison with equation (3.6) in \cite{ettle_and_morris}, they reduce to
\begin{equation}	\nonumber\sum_{k=i}^{j-1}\frac{\left\{k,k+1\right\}}{\hat{k}\widehat{k+1}}\left(q,P_{i,k}\right)=-\hat{q}\left(\omega_{-q}+\omega_i+\cdots+\omega_j\right)
\end{equation}
with $q=p_i+\cdots+p_j=-p_{j+1}-\cdots-p_n-p_1+\cdots-p_{i-1}$ and $\omega_p=\bar{p}\tilde{p}/\hat{p}$. So we have

\begin{equation}\begin{split} \nonumber Y_{i,j}=\frac{\hat{1}^2}{\hat{q}^2}\Big(\hat{q}^G\Big\{\omega_{-q}^G+\omega_i^G+\cdots+\omega_j^G\Big\}+\hat{q}\Big\{\omega_{q}+\omega_{j+1}+\cdots+\omega_{i-1}\Big\}\Big)\times\\
	\times\Gamma\left(q^G,i^G,\cdots,j^G\right)\Gamma\left(q,j+1,\cdots,i-1\right).
\end{split}\end{equation} 
Now since $-q+p_i+\cdots+p_j=0$ and $q+p_{j+1}+\cdots+p_n+p_1+\cdots+p_{i-1}=0$ we can subtract these from each of the brackets $\omega_{-q}+p_i+\cdots+p_j$ as follows

\begin{equation}\begin{split}	 \nonumber Y_{i,j}=\frac{\hat{1}^2}{\hat{q}^2}\Bigg(\hat{q}^G\Big\{\omega_{-q}^G-\widecheck{-q}^G+\omega_i^G-\check{i}^G&+\cdots+\omega_j^G-\check{j}^G\Big\}+\\	&+\hat{q}\Big\{\omega_{q}-\check{q}+\omega_{j+1}-\widecheck{j+1}+\cdots+\omega_{i-1}-\widecheck{i-1}\Big\}\Bigg)\times\\
&\times\Gamma\left(q^G,i^G,\cdots,j^G\right)\Gamma\left(q,j+1,\cdots,i-1\right)
\end{split}\end{equation}
and then take out a factor of $1/\hat{p}$ from each term $\omega_P-\check{p}$ as follows
\begin{eqnarray}
	\nonumber	Y_{i,j}=\frac{\hat{1}^2}{\hat{q}^2}\left(\frac{\hat{q}^G}{\hat{q}^G}\Omega_{q}^G+\frac{\hat{q}^G}{\hat{i}^G}\Omega_i^G+\cdots+\frac{\hat{q}^G}{\hat{j}^G}\Omega_j^G-\frac{\hat{q}}{\hat{q}}\Omega_{-q}+\frac{\hat{q}}{\widehat{j+1}}\Omega_{j+1}+\cdots+\frac{\hat{q}}{\widehat{i-1}}\Omega_{i-1}\right)\times\\
\nonumber\times\Gamma\left(q^G,i^G,\cdots,j^G\right)\Gamma\left(q,j+1,\cdots,i-1\right).
\end{eqnarray}
Terms in $\Omega_q$ and $\Omega_{-q}$ cancel, using the fact that $\Omega^G=\Omega$. So we arrive at
\begin{eqnarray}
	\nonumber	Y_{i,j}=\frac{\hat{1}^2}{\hat{q}^2}\left(\frac{\hat{q}}{\hat{1}}\Omega_1+\cdots+\frac{\hat{q}}{\hat{i-1}}\Omega_{i-1}+\frac{\hat{q}^G}{\hat{i}^G}\Omega_i^G+\cdots+\frac{\hat{q}^G}{\hat{j}^G}\Omega_j^G+\frac{\hat{q}}{\widehat{j+1}}\Omega_{j+1}+\cdots+\frac{\hat{q}}{\hat{n}}\Omega_n\right)\times\\	\nonumber\times\Gamma\left(q^G,i^G,\cdots,j^G\right)\Gamma\left(q,j+1,\cdots,i-1\right)\\
	\nonumber=-X_{i,j}
\end{eqnarray}
and so coefficients of linearly independent, like terms in $\overline{A}_{\bar{1}}\cdots A_{\bar{i}^G}\cdots A_{\bar{j}^G}\cdots A_{\bar{n}}$ in $\delta S_F$ and $\delta S_I$ sum to zero, $X_{i,j}+Y_{i,j}=0$ and the result follows
\begin{eqnarray}
\nonumber\delta S=\delta S_F+\delta S_I=0
\end{eqnarray}
thus not only proving that the expressions (\ref{eq:deltaA}) and (\ref{eq:deltaAbar}) and their associated momentum flow diagrams are indeed symmetries of the Chalmers-Siegel action and the most general transformations in fig (3) are also symmetries of the action (\ref{eq:momentum space action}).

\section{Conclusion and Summary}
The Chalmers-Siegel action which describes self-dual Yang-Mills theory can be mapped to a free theory by a canonical transformation arising from the construction of a lagrangian formalism of the MHV rules. Free theories have a high degree of symmetry. In addition to the well-known symmetries induced by infinitesimal isometries there are those in which infinitesimal changes in the fields are related to finite isometries which we have reviewed briefly. The Lie algebra of these transformations is built out of the group algebra of the isometries, and this can be used to decompose the Lie algebra into a direct sum of its Abelian and non-Abelian parts. By studying the canonical transformation we found the corresponding symmetries of the self-dual Yang-Mills theory, and showed that these satisfy the same Lie algebra as in the free theory. We expect that these results are generalisable to the supersymmetric case and in particular to $N=4$ super Yang-Mills on the light cone. It will also be interesting to see which (if any) of these symmetries survive the full Yang-Mills theory on the light cone, given by eqn (\ref{eq:fulllightconeaction}). We expect only a subset of the transformations to survive. Further, by considering the dihedral subgroups $D(2n)$ of $SO(3)$ and counting the number af Abelian generators, we find that the number of non-Abelian generators increases in multiples of $3$ with increasing $n$. We expect to find that the algebra constructed in this way using the dihedral groups is going to be a sum of $su(2)$ algebras.

Throughout this paper we have restricted ourselves to studying the isometries of the Lorentz group. Extending the result to the include displacements is somewhat more trivial and a phase factor appears in the expressions using the fundamental property of Fourier transforms that
\begin{equation}
\nonumber\phi(x+a)\stackrel{FT}{\rightarrow} e^{ipa}\tilde{\phi}(p).
\end{equation}
For example under a pure translation $x\rightarrow x+a$, following the same procedure to the one we have given throughout, we would get
\begin{eqnarray}
\label{eq:deltaA}
	\nonumber\delta A_1=-\epsilon\sum_{n=2}^{\infty}\sum_{i=2}^{n}\sum_{j=i}^n\int_{2\cdots n}\frac{\hat{1}}{\hat{q}}\Gamma(q,i,\cdots,j)\Gamma(q,j+1,\cdots,n,1\cdots,i-1)\times\\
\nonumber	\times exp\left\{i(p_{\mu}^i+\cdots+p_{\mu}^j)a^{\mu}\right\}A_{\bar{2}}\cdots A_{\bar{n}}
\end{eqnarray}
neglecting various factors of $2\pi$. A similar expression would hold for the transformation of the conjugate field $\overline{\delta{A}}$ with the exponential factors appearing in each term of the sum.

\acknowledgments
PM thanks STFC for support under the rolling grant ST/G000433/1 and AW
thanks STFC for a studentship.
\appendix
\allowdisplaybreaks
\section{Order by Order Calculation of $\delta A$}
\label{ap:A}
Expanding $\delta A$ in terms of $B$ as per (\ref{eq:delA in terms of B}) and substituting $B[A]$ up to fourth order, we arrive at the following expression
\begin{eqnarray}
\label{eq:delta A expansion}
	\nonumber \delta A_{\bar{1}}=\epsilon A_{\bar{1}^G}&+&\int_{23}\Gamma(123)A_{\bar{2}^G}A_{\bar{3}^G}+\int_{234}\Gamma(1234)A_{\bar{2}^G}A_{\bar{3}^G}A_{\bar{4}^G}\\
\nonumber	&+&\int_{2345}\Gamma(12345)A_{\bar{2}^G}A_{\bar{3}^G}A_{\bar{4}^G}A_{\bar{5}^G}\\
\nonumber\\
\nonumber &+&\epsilon\int_{23}\Upsilon(123)\left(A_{\bar{2}^G}+\int_{45}\Gamma(\bar{2}45)A_{\bar{4}^G}A_{\bar{5}^G}+\int_{456}\Gamma(\bar{2}456)A_{\bar{4}^G}A_{\bar{5}^G}A_{\bar{6}^G}\right)\times \\
\nonumber &&\ \ \ \ \ \ \ \ \ \ \ \ \ \ \times \left(A_{\bar{3}}+\int_{78}\Gamma(\bar{3}78)A_{\bar{7}}A_{\bar{8}}+\int_{789}\Gamma(\bar{3}789)A_{\bar{7}}A_{\bar{8}}A_{\bar{9}}\right)\\
\nonumber\\
\nonumber &+&\epsilon\int_{23}\Upsilon(123)\left(A_{\bar{2}}+\int_{45}\Gamma(\bar{2}45)A_{\bar{4}}A_{\bar{5}}+\int_{456}\Gamma(\bar{2}456)A_{\bar{4}}A_{\bar{5}}A_{\bar{6}}\right)\times\\
\nonumber &&\ \ \ \ \ \ \ \ \ \ \ \ \ \  \times\left(A_{\bar{3}^G}+\int_{78}\Gamma(\bar{3}78)A_{\bar{7}^G}A_{\bar{8}^G}+\int_{789}\Gamma(\bar{3}789)A_{\bar{7}^G}A_{\bar{8}^G}A_{\bar{9}^G}\right)\\
\nonumber\\
\nonumber &+&\epsilon\int_{234}\Upsilon(1234)\left(A_{\bar{2}^G}+\int_{56}\Gamma(\bar{2}56)A_{\bar{5}^G}A_{\bar{6}^G}\right)\left(A_{\bar{3}}+\int_{78}\Gamma(\bar{3}78)A_{\bar{7}}A_{\bar{8}}\right)\times\\
\nonumber&& \ \ \ \ \ \ \ \ \ \ \ \ \ \ \times\left(A_{\bar{4}}+\int_{9\;10}\Gamma(\bar{4}9\;10)A_{\bar{9}}A_{\bar{10}}\right)\\
\nonumber\\
\nonumber &+&\epsilon\int_{234}\Upsilon(1234)\left(A_{\bar{2}}+\int_{56}\Gamma(\bar{2}56)A_{\bar{5}}A_{\bar{6}}\right)\left(A_{\bar{3}^G}+\int_{78}\Gamma(\bar{3}78)A_{\bar{7}^G}A_{\bar{8}^G}\right)\times\\
\nonumber&&\ \ \ \ \ \ \ \ \ \ \ \ \ \ \ \times\left(A_{\bar{4}}+\int_{9\;10}\Gamma(\bar{4}9\;10)A_{\bar{9}}A_{\bar{10}}\right)\\
\nonumber\\
\nonumber &+&\epsilon\int_{234}\Upsilon(1234)\left(A_{\bar{2}}+\int_{56}\Gamma(\bar{2}56)A_{\bar{5}}A_{\bar{6}}\right)\left(A_{\bar{3}}+\int_{78}\Gamma(\bar{3}78)A_{\bar{7}}A_{\bar{8}}\right)\times\\
\nonumber&&\ \ \ \ \ \ \ \ \ \ \ \ \ \ \ \ \times\left(A_{\bar{4}^G}+\int_{9\;10}\Gamma(\bar{4}9\;10)A_{\bar{9}^G}A_{\bar{10}^G}\right)\\
\nonumber\\
\nonumber &+&\epsilon\int_{2345}\Upsilon(12345)A_{\bar{2}^G}A_{\bar{3}}A_{\bar{4}}A_{\bar{5}}+\epsilon\int_{2345}\Upsilon(12345)A_{\bar{2}}A_{\bar{3}^G}A_{\bar{4}}A_{\bar{5}}\\
\nonumber &+&\epsilon\int_{2345}\Upsilon(12345)A_{\bar{2}}A_{\bar{3}}A_{\bar{4}^G}A_{\bar{5}}+\epsilon\int_{2345}\Upsilon(12345)A_{\bar{2}}A_{\bar{3}}A_{\bar{4}}A_{\bar{5}^G}.
\end{eqnarray}
Despite looking horrendous, when like terms are collected and their coefficients calculated the expression simplifies into something more tangeable. We shall collect terms order by order. The reader may wish to study an example to lower orders first, say cubic terms to become used to the calculations. First order is trivial, we get $\delta A=\epsilon A_{1^G}+\cdots$.

Second order isn't much more difficult, we simply find the terms that are quadratic in the A fields when expanding out the brackets. We get
\begin{equation}
	\nonumber\delta A_{1}=\epsilon A_{1^G}+\epsilon\int_{23}\left\{\Gamma(123)A_{\bar{2}^G}A_{\bar{3}^G}+\Upsilon(123)A_{\bar{2}^G}A_{\bar{3}}+\Upsilon(123)A_{\bar{2}}A_{\bar{3}^G}\right\}+\cdots.
\end{equation}
Further, when $\Gamma$ and $\Upsilon$ are expressed in terms of their independent momenta the expression is
\begin{equation}
\nonumber	\delta A_{1}=\epsilon A_{1^G}+\epsilon\int_{23}\left\{i\frac{\hat{1}}{(23)}A_{\bar{2}^G}A_{\bar{3}^G}-i\frac{\hat{1}}{(23)}A_{\bar{2}^G}A_{\bar{3}}-i\frac{\hat{1}}{(23)}A_{\bar{2}}A_{\bar{3}^G}\right\}+\cdots.
\end{equation}
Third order gets more tricky. Taking the third order terms out of the expansion, we get
\begin{eqnarray}
	\nonumber \cdots &+&\epsilon\int_{234}\Gamma(1234)A_{\bar{2}^G}A_{\bar{3}^G}A_{\bar{4}^G}+\\	
	\nonumber &+&\epsilon\int_{2378}\Upsilon(123)\Gamma(\bar{3}78)A_{\bar{2}^G}A_{\bar{7}^G}A_{\bar{8}}+\epsilon\int_{2345}\Upsilon(123)\Gamma(\bar{2}45)A_{\bar{4}}A_{\bar{5}^G}A_{\bar{3}^G}\\	\nonumber &+&\epsilon\int_{2378}\Upsilon(123)\Gamma(\bar{3}78)A_{\bar{2}}A_{\bar{7}^G}A_{\bar{8}^G}+\epsilon\int_{2345}\Upsilon(123)\Gamma(\bar{2}45)A_{\bar{4}^G}A_{\bar{5}^G}A_{\bar{3}}\\	\nonumber&+&\epsilon\int_{234}\Upsilon(1234)A_{\bar{2}^G}A_{\bar{3}}A_{\bar{4}}+\epsilon\int_{234}\Upsilon(1234)A_{\bar{2}}A_{\bar{3}^G}A_{\bar{4}}\\
\nonumber	&+&\epsilon\int_{234}\Upsilon(1234)A_{\bar{2}}A_{\bar{3}}A_{\bar{4}^G}+\cdots.
\end{eqnarray}
Now we carefully change variables of integration, maintaining the order of the fields since they contain group matrices, and collect terms,
\begin{equation}\begin{split}	\nonumber\cdots+\epsilon\int_{234}\bigg\{&\Gamma(1234)A_{\bar{2}^G}A_{\bar{3}^G}A_{\bar{4}^G}+\Upsilon(154)\Gamma(\bar{5}23)A_{\bar{2}^G}A_{\bar{3}^G}A_{\bar{4}}+\Upsilon(125)\Gamma(\bar{5}34)A_{\bar{2}}A_{\bar{3}}A_{\bar{4}^G}+\\
	&+\left\{\Upsilon(125)\Gamma(\bar{5}34)+\Upsilon(1234)\right\}A_{\bar{2}^G}A_{\bar{3}}A_{\bar{4}}+\left\{\Upsilon(154)\Gamma(\bar{5}23)+\Upsilon(1234)\right\}A_{\bar{2}}A_{\bar{3}}A_{\bar{4}^G}+\\
	&+\Upsilon(1234)A_{\bar{2}}A_{\bar{3}^G}A_{\bar{4}}\bigg\}+\cdots
\end{split}\end{equation}
where $p_5$ is minus the sum of the remaining arguments in the coefficient. For example, in the second term, $p_5=-p_1-p_4=+p_2+p_3$. Remarkably, when expressed in terms of their independent momenta they reduce to simpler expressions, in particular the fourth and fifth terms whose coefficients are $\left\{\Upsilon(125)\Gamma(\bar{5}23)+\Upsilon(1234)\right\}$ and $\left\{\Upsilon(154)\Gamma(\bar{5}23)+\Upsilon(1234)\right\}$ respectively reduce nicely. For example, take the fifth coefficient bearing in mind momentum conservation, $1+2+3+4=0$,
\begin{equation}	\nonumber\Upsilon(154)\Gamma(\bar{5}23)+\Upsilon(1234)=\left(-i\frac{\hat{1}}{(25)}\right)\left(i\frac{\hat{\bar{5}}}{(34)}\right)+\frac{\hat{1}}{(23)}\frac{\hat{3}}{(34)}
\end{equation}
taking out a factor $\hat{1}/(34)$ gives
\begin{equation}
\nonumber	\frac{\hat{1}}{(34)}\left(\frac{\hat{1}+\hat{2}}{(12)}+\frac{\hat{3}}{(23)}\right)
\end{equation}
then putting the expression in brackets under a common denominator and expanding out terms on the numerator
\begin{equation}
\nonumber\frac{\hat{1}}{(34)}\left(\frac{\hat{1}\hat{2}\tilde{3}+\hat{2}\hat{2}\tilde{3}-\hat{2}\tilde{2}\hat{3}-\hat{3}\tilde{1}\hat{2}}{(12)(23)}\right)	
\end{equation}
giving
\begin{equation}
\nonumber	\frac{\hat{1}\hat{2}}{(12)(23)}=\frac{\hat{1}\hat{2}}{(23)(2,3+4)}
\end{equation}
expressing the coefficients in this way and using momentum conservation to express the denominators in a certain way, the third order expression is
\begin{equation}\begin{split}
\nonumber \cdots+\epsilon\int_{234}\bigg\{&\frac{\hat{1}\hat{q}\ A_{\bar{2}^G}A_{\bar{3}^G}A_{\bar{4}^G}}{(q,2)(q,2+3)}+\frac{\hat{1}\hat{q}\ A_{\bar{2}^G}A_{\bar{3}^G}A_{\bar{4}}}{(q,2)(q,4)}+\frac{\hat{1}\hat{q}\ A_{\bar{2}}A_{\bar{3}^G}A_{\bar{4}^G}}{(q,3)(q,1)}\\
&+\frac{\hat{1}\hat{q}\ A_{\bar{2}^G}A_{\bar{3}}A_{\bar{4}}}{(q,3)(q,3+4)}+\frac{\hat{1}\hat{q}\ A_{\bar{2}}A_{\bar{3}^G}A_{\bar{4}}}{(q,4)(q,4+1)}+\frac{\hat{1}\hat{q}\ A_{\bar{2}}A_{\bar{3}}A_{\bar{4}^G}}{(q,1)(q,1+2)}\bigg\}+\cdots
\end{split}\end{equation}
where for any term with $A_{\bar{2}}\cdots A_{\bar{i}^G} \cdots A_{\bar{j}^G} \cdots A_{\bar{n}}$, $q$ is defined to be $q=p_i+ \cdots +p_j$. Given these expressions it is tempting to substitute $q=p_i+ \cdots +p_j$ and simplify the coefficients further. However, we write the terms like this deliberately because as we shall see fourth order terms follow a similar pattern which would not otherwise be visible. 

We can collect together terms that are quartic in A within the confines of an A4 page too. Doing so, and carefully relabelling variables of integration, we arrive at

\begin{equation}\begin{split}
\label{eq:dA4thorder}
	\nonumber\cdots+\epsilon\int_{2345}\bigg\{&\Gamma(12345)A_{\bar{2}^G}A_{\bar{3}^G}A_{\bar{4}^G}A_{\bar{5}^G}+\\	&+\Upsilon(165)\Gamma(\bar{6}234)A_{\bar{2}^G}A_{\bar{3}^G}A_{\bar{4}^G}A_{\bar{5}}+\Upsilon(126)\Gamma(\bar{6}345)A_{\bar{2}}A_{\bar{3}^G}A_{\bar{4}^G}A_{\bar{5}^G}+\\	&+\left\{\Upsilon(167)\Gamma(\bar{6}23)\Gamma(\bar{7}45)+\Upsilon(1645)\Gamma(\bar{6}23)\right\}A_{\bar{2}^G}A_{\bar{3}^G}A_{\bar{4}}A_{\bar{5}}+\\
&+\Upsilon(1265)\Gamma(\bar{6}34)A_{\bar{2}}A_{\bar{3}^G}A_{\bar{4}^G}A_{\bar{5}}+\\	&+\left\{\Upsilon(167)\Gamma(\bar{6}23)\Gamma(\bar{7}45)+\Upsilon(1236)\Gamma(\bar{6}45)\right\}A_{\bar{2}}A_{\bar{3}}A_{\bar{4}^G}A_{\bar{5}^G}+\\	&+\left\{\Upsilon(126)\Gamma(\bar{6}345)+\Upsilon(1236)\Gamma(\bar{6}45)+\Upsilon(1265)\Gamma(\bar{6}34)+\Upsilon(12345)\right\}A_{\bar{2}^G}A_{\bar{3}}A_{\bar{4}}A_{\bar{5}}+\\
&+\left\{\Upsilon(1236)\Gamma(\bar{6}45)+\Upsilon(12345)\right\}A_{\bar{2}}A_{\bar{3}^G}A_{\bar{4}}A_{\bar{5}}+\\
&+\left\{\Upsilon(1645)\Gamma(\bar{6}23)+\Upsilon(12345)\right\}A_{\bar{2}}A_{\bar{3}}A_{\bar{4}^G}A_{\bar{5}}+\\	&+\left\{\Upsilon(165)\Upsilon(\bar{6}234)+\Upsilon(1265)\Gamma(\bar{6}34)+\Upsilon(1645)\Gamma(\bar{6}23)+\Upsilon(12345)\right\}A_{\bar{2}}A_{\bar{3}}A_{\bar{4}}A_{\bar{5}^G}\bigg\}+\cdots.\\
\end{split}\end{equation}
The above expression simplifies in a similar way to earlier. We shall state the result first, and give an example of one of the calculations. The others are similar, the most complicated ones can be checked on a computer algebra package.
\begin{equation}\begin{split}	\cdots+\epsilon\int_{2345}i\bigg\{&\frac{\hat{1}\hat{q}^2A_{\bar{2}^G}A_{\bar{3}^G}A_{\bar{4}^G}A_{\bar{5}^G}}{(q,2)(q,2+3)(q,2+3+4)}A_{\bar{2}^G}+\frac{\hat{1}\hat{q}^2A_{\bar{2}^G}A_{\bar{3}^G}A_{\bar{4}^G}A_{\bar{5}}}{(q,2)(q,2+3)(q,5)}+\\
&+\frac{\hat{1}\hat{q}^2A_{\bar{2}}A_{\bar{3}^G}A_{\bar{4}^G}A_{\bar{5}^G}}{(q,3)(q,3+4)(q,1)}+\frac{\hat{1}\hat{q}^2A_{\bar{2}^G}A_{\bar{3}^G}A_{\bar{4}}A_{\bar{5}}}{(q,2)(q,4)(q,4+5)}+\\
&+\frac{\hat{1}\hat{q}^2A_{\bar{2}}A_{\bar{3}^G}A_{\bar{4}^G}A_{\bar{5}}}{(q,3)(q,5)(q,5+1)}+\frac{\hat{1}\hat{q}^2A_{\bar{2}}A_{\bar{3}}A_{\bar{4}^G}A_{\bar{5}^G}}{(q,4)(q,1)(q,1+2)}\\	&+\frac{\hat{1}\hat{q}^2A_{\bar{2}^G}A_{\bar{3}}A_{\bar{4}}A_{\bar{5}}}{(q,3)(q,3+4)(q,3+4+5)}+\frac{\hat{1}\hat{q}^2A_{\bar{2}}A_{\bar{3}^G}A_{\bar{4}}A_{\bar{5}}}{(q,4)(q,4+5)(q,4+5+1)}+\\	
&+\frac{\hat{1}\hat{q}^2A_{\bar{2}}A_{\bar{3}}A_{\bar{4}^G}A_{\bar{5}}}{(q,5)(q,5+1)(q,5+1+2)}+\frac{\hat{1}\hat{q}^2A_{\bar{2}}A_{\bar{3}}A_{\bar{4}}A_{\bar{5}^G}}{(q,1)(q,1+2)(q,1+2+3)}\bigg\}+\cdots.
\end{split}\end{equation}
As an example, let us take the fourth term
\begin{eqnarray}	\nonumber\Upsilon(167)\Gamma(\bar{6}23)\Gamma(\bar{7}45)+\Upsilon(1645)\Gamma(\bar{6}23)=i\frac{\hat{1}(\hat{2}+\hat{3})}{(23)(45)}\left\{\frac{\hat{4}+\hat{5}}{(2+3,4+5)}-\frac{\hat{4}}{(2+3,4)}\right\},
\end{eqnarray}
upon expressing the coefficients $\Gamma$ and $\Upsilon$ explicitly in terms of their independent momenta and taking out a factor $i\frac{\hat{1}(\hat{2}+\hat{3})}{(23)(45)}$. Further, we substitute $p_4+p_5=-p_1-p_2-p_3$ and take out a factor of $-1$
\begin{equation}	\nonumber=-i\frac{\hat{1}(\hat{2}+\hat{3})}{(23)(45)}\left\{\frac{\hat{1}+\hat{2}+\hat{3}}{(2+3,4+5)}+\frac{\hat{4}}{(2+3,4)}\right\}.
\end{equation}
Let's call $q=2+3$, and put the term in brackets under a common denominator i.e,
\begin{equation}
\nonumber	=-i\frac{\hat{1}\hat{q}}{(23)(45)}\left(\frac{(\hat{1}+\hat{q})(q,4)+\hat{4}(1,q)}{(1,x)(q,4)}\right).
\end{equation}
Now expand the numerator, two terms cancel, then a factor of $\hat{q}$ can be taken outside the bracket giving
\begin{equation}
\nonumber	=-i\frac{\hat{1}\hat{q}^2}{(23)(45)(1,q)(q,4)}(1+q,4)=-i\frac{\hat{1}(\hat{2}+\hat{3})^2}{(23)(1,2+3)(2+3,4)}(45).
\end{equation}
Then using momentum conservation, we get the simplified coefficient
\begin{equation}
\nonumber	\Upsilon(167)\Gamma(\bar{6}23)\Gamma(\bar{7}45)+\Upsilon(1645)\Gamma(\bar{6}23)=i\frac{\hat{1}\hat{q}^2}{(q,2)(q,4)(q,4+5)}
\end{equation}
where $q=p_2+p_3$. Calculation of the other terms is equally as simple.
\section{Order by Order Calculation of $\delta \overline{A}$}
\label{ap:B}
We expand $\delta\overline{A}$ in terms of the free field $B$, $\overline{B}$, $\delta B$ and $\delta\overline{B}$ and as per  (\ref{eq:delAbar in terms of B}). working to third order only. In a similar fashion to the calculation of appendix (\ref{ap:A}), we substitute the inverse expressions, $B[A]$ and  $\overline{B}[A,\overline{A}]$ which is given by the expansion
\begin{eqnarray}	\nonumber\overline{B}_{\bar{1}}=\overline{A}_{\bar{1}}+\int_{23}&\bigg\{&\frac{\hat{2}}{\hat{1}}\Theta^2(\bar{1}23)\overline{A}_{\bar{2}}A_{\bar{3}}+\frac{\hat{3}}{\hat{1}}\Theta^3(\bar{1}23)A_{\bar{2}}\overline{A}_{\bar{3}}+\\
\nonumber&+&\frac{\hat{2}}{\hat{1}}\Theta^2(\bar{1}234)\overline{A}_{\bar{2}}A_{\bar{3}}A_{\bar{4}}+\frac{\hat{3}}{\hat{1}} \Theta^3(\bar{1}234)A_{\bar{2}}\overline{A}_{\bar{3}}A_{\bar{4}}+\frac{\hat{4}}{\hat{1}}\Theta^4(\bar{1}234)A_{\bar{2}}A_{\bar{3}}\overline{A}_{\bar{4}}\bigg\}+\cdots.
\end{eqnarray}
Performing this substitution, maintaining third order terms only, we arrive at
\begin{eqnarray}	\nonumber\delta\overline{A}_1&=&-\epsilon\overline{A}_{1^{G^{-1}}}+\epsilon\int_{23}\frac{\hat{2}}{\hat{1}}\Theta^2(123)\overline{A}_{\bar{2}^{G^{-1}}}A_{\bar{3}^{G^{-1}}}+\epsilon\int_{23}\frac{\hat{3}}{\hat{1}}\Theta^3(123)A_{\bar{2}^{G^{-1}}}\overline{A}_{\bar{3}^{G^{-1}}}+\\	\nonumber&+&\epsilon\int_{234}\frac{\hat{4}}{\hat{1}}\Theta^4(1234)\overline{A}_{\bar{2}^{G^{-1}}}A_{\bar{3}^{G^{-1}}}A_{\bar{4}^{G^{-1}}}+\epsilon\int_{234}\frac{\hat{3}}{\hat{1}}\Theta^3(1234)A_{\bar{2}^{G^{-1}}}\overline{A}_{\bar{3}^{G^{-1}}}A_{\bar{4}^{G^{-1}}}\\
\nonumber&+&\epsilon\int_{234}\frac{\hat{2}}{\hat{1}}\Theta^2(1234)\overline{A}_{\bar{2}^{G^{-1}}}A_{\bar{3}^{G^{-1}}}A_{\bar{4}^{G^{-1}}}\\
\nonumber\\
\nonumber&+&\epsilon\int_{23}\frac{\hat{2}}{\hat{1}}\Xi^2(123)\left(\overline{A}_{\bar{2}^{G^{-1}}}+\int_{45}\frac{\hat{4}}{\hat{2}}\Theta^2(\bar{2}45)\overline{A}_{\bar{4}^{G^{-1}}}A_{\bar{5}^{G^{-1}}}+\int_{45}\frac{\hat{5}}{\hat{2}}\Theta^3(\bar{2}45)A_{\bar{4}^{G^{-1}}}\overline{A}_{\bar{5}^{G^{-1}}}\right)\times\\
\nonumber&&\ \ \ \ \ \ \ \ \ \ \ \ \ \times\left(A_{\bar{3}}+\int_{67}\Gamma(\bar{3}67)A_{\bar{6}}A_{\bar{7}}\right)\\
\nonumber\\
\nonumber&-&\epsilon\int_{23}\frac{\hat{2}}{\hat{1}}\Xi^2(123)\left(\overline{A}_{\bar{2}}+\int_{45}\frac{\hat{4}}{\hat{2}}\Theta^2(\bar{2}45)\overline{A}_{\bar{4}}A_{\bar{5}}+\int_{45}\frac{\hat{5}}{\hat{2}}\Theta^3(\bar{2}45)A_{\bar{4}}\overline{A}_{\bar{5}}\right)\times\\
\nonumber&&\ \ \ \ \ \ \ \ \ \ \ \ \ \ \ \times\left(A_{\bar{3}^G}+\int_{67}\Gamma(\bar{3}67)A_{\bar{6}^G}A_{\bar{7}^G}\right)\\
\nonumber\\
\nonumber&-&\epsilon\int_{23}\frac{\hat{3}}{\hat{1}}\Xi^3(123)\left(A_{\bar{2}^G}+\int_{45}\Gamma(\bar{2}45)A_{\bar{4}^G}A_{\bar{5}}\right)\times\\
\nonumber&&\ \ \ \ \ \ \ \ \ \ \ \ \ \ \ \times\left(\overline{A}_{\bar{3}}+\int_{67}\frac{\hat{6}}{\hat{3}}\Theta^2(\bar{3}67)\overline{A}_{\bar{6}}A_{\bar{7}}+\int_{67}\frac{\hat{7}}{\hat{3}}\Theta^3(\bar{3}67)A_{\bar{6}}\overline{A}_{\bar{7}}\right)\\
\nonumber\\
\nonumber&+&\epsilon\int_{23}\frac{\hat{3}}{\hat{1}}\Theta^3(123)\left(A_{\bar{2}}+\int_{45}\Gamma(\bar{2}45)A_{\bar{4}}A_{\bar{5}}\right)\times\\
\nonumber&&\ \ \ \ \ \ \ \ \ \ \ \ \ \ \ \times\left(\overline{A}_{\bar{3}^{G^{-1}}}+\int_{67}\frac{\hat{6}}{\hat{3}}\Xi^2(\bar{3}67)\overline{A}_{\bar{6}^{G^{-1}}}A_{\bar{7}^{G^{-1}}}+\int_{67}\frac{\hat{7}}{\hat{3}}\Xi^3(\bar{3}67)\overline{A}_{\bar{6}^{G^{-1}}}A_{\bar{7}^{G^{-1}}}\right)\\
\nonumber\\
\nonumber&+&\epsilon\int_{234}\frac{\hat{2}}{\hat{1}}\Xi^2(1234)\overline{A}_{\bar{2}^{G^{-1}}}A_{\bar{3}}A_{\bar{4}}-\epsilon\int_{234}\frac{\hat{2}}{\hat{1}}\Xi^2(1234)\overline{A}_{\bar{2}}A_{\bar{3}^G}A_{\bar{4}}-\epsilon\int_{234}\frac{\hat{2}}{\hat{1}}\Xi^2(1234)\overline{A}_{\bar{2}}A_{\bar{3}}A_{\bar{4}^G}\\
\nonumber\\
\nonumber&-&\epsilon\int_{234}\frac{\hat{3}}{\hat{1}}\Xi^3(1234)A_{\bar{2}^G}\overline{A}_{\bar{3}}A_{\bar{4}}+\epsilon\int_{234}\frac{\hat{3}}{\hat{1}}\Xi^3(1234)A_{\bar{2}}\overline{A}_{\bar{3}^{G^{-1}}}A_{\bar{4}}-\epsilon\int_{234}\frac{\hat{3}}{\hat{1}}\Xi^3(1234)A_{\bar{2}}\overline{A}_{\bar{3}}A_{\bar{4}^G}\\
\nonumber\\
\nonumber&-&\epsilon\int_{234}\frac{\hat{4}}{\hat{1}}\Xi^4(1234)A_{\bar{2}^G}A_{\bar{3}}\overline{A}_{\bar{4}}-\epsilon\int_{234}\frac{\hat{4}}{\hat{1}}\Xi^4(1234)A_{\bar{2}}A_{\bar{3}^G}\overline{A}_{\bar{4}}+\epsilon\int_{234}\frac{\hat{4}}{\hat{1}}\Xi^4(1234)A_{\bar{2}}A_{\bar{3}}\overline{A}_{\bar{4}^{G^{-1}}}+\cdots.
\end{eqnarray}
Again, we shall collect terms order by order we shall see that we have already done most of the work already when calculating the coefficients in appendix (\ref{ap:A}). First order is again trivial, we get $\delta \overline{A}_1=-\epsilon\overline{A}_{1^{G^-1}}+\cdots$. At second order we can pick out the terms and express $\Xi$ and $\Theta$ in terms of independent momenta, no extra calculation is required.
\begin{eqnarray}	\nonumber\delta\overline{A}_1=-\epsilon\overline{A}_{1^{G^{-1}}}+\epsilon\int_{23}&\bigg\{&\frac{\hat{2}}{\hat{1}}\Theta^2(123)\overline{A}_{\bar{2}^{G^{-1}}}A_{\bar{3}^{G^{-1}}}+\frac{\hat{3}}{\hat{1}}\Theta^3(123)A_{\bar{2}^{G^{-1}}}\overline{A}_{\bar{3}^{G^{-1}}}\\
\nonumber&+&\frac{\hat{2}}{\hat{1}}\Xi^2(123)\overline{A}_{\bar{2}^{G^{-1}}}A_{\bar{3}}-\frac{\hat{2}}{\hat{1}}\Xi^2(123)\overline{A}_{\bar{2}}A_{\bar{3}^G}\\
\nonumber&-&\frac{\hat{3}}{\hat{1}}\Xi^3(123)A_{\bar{2}^G}\overline{A}_{\bar{3}}+\frac{\hat{3}}{\hat{1}}\Xi^3(123)A_{\bar{2}}\overline{A}_{\bar{3}^{G^{-1}}}\bigg\}+\cdots.
\end{eqnarray}
As per \cite{ettle_and_morris} we have
\begin{eqnarray}
	\nonumber\Xi^2(123)=-\Upsilon(231)=i\frac{\hat{2}}{(31)}\\
\nonumber	\Xi^3(123)=-\Upsilon(312)=i\frac{\hat{3}}{(12)}
\end{eqnarray}
and we can deduce for ourselves
\begin{eqnarray}
	\nonumber\Theta^2(123)&=&-\Gamma(231)=-i\frac{\hat{2}}{(31)}\\
	\nonumber\Theta^3(123)&=&-\Gamma(312)=-i\frac{\hat{3}}{(12)}
\end{eqnarray}
so to second order we find
\begin{eqnarray}	\nonumber\delta\overline{A}_1=-\epsilon\overline{A}_{1^{G^{-1}}}-\epsilon\int_{23}i&\bigg\{&\frac{\hat{2}}{\hat{1}}\frac{\hat{2}}{(31)}\overline{A}_{\bar{2}^{G^{-1}}}A_{\bar{3}^{G^{-1}}}-\frac{\hat{3}}{\hat{1}}\frac{\hat{3}}{(12)}A_{\bar{2}^{G^{-1}}}\overline{A}_{\bar{3}^{G^{-1}}}\\
\nonumber&+&\frac{\hat{2}}{\hat{1}}\frac{\hat{2}}{(31)}\overline{A}_{\bar{2}^{G^{-1}}}A_{\bar{3}}-\frac{\hat{2}}{\hat{1}}\frac{\hat{2}}{(31)}\overline{A}_{\bar{2}}A_{\bar{3}^G}\\
\nonumber &-&\frac{\hat{3}}{\hat{1}}\frac{\hat{3}}{(12)}A_{\bar{2}^G}\overline{A}_{\bar{3}}+\frac{\hat{3}}{\hat{1}}\frac{\hat{3}}{(12)}A_{\bar{2}}\overline{A}_{\bar{3}^{G^{-1}}}\bigg\}+\cdots.
\end{eqnarray}
Finally to third order, careful inspection of the expansion will produce the following, where we have carefully relabelled variables,
\begin{eqnarray}	\nonumber\cdots&+&\epsilon\int_{234}\bigg\{\frac{\hat{2}}{\hat{1}}\Theta^2(1234)\overline{A}_{\bar{2}^{G^{-1}}}A_{\bar{3}^{G^{-1}}}A_{\bar{4}^{G^{-1}}}+\frac{\hat{3}}{\hat{1}}\Theta^3(1234)A_{\bar{2}^{G^{-1}}}\overline{A}_{\bar{3}^{G^{-1}}}A_{\bar{4}^{G^{-1}}}\\	\nonumber&+&\frac{\hat{4}}{\hat{1}}\Theta^4(1234)A_{\bar{2}^{G^{-1}}}A_{\bar{3}^{G^{-1}}}\overline{A}_{\bar{4}^{G^{-1}}}\\	\nonumber&+&\frac{\hat{2}}{\hat{1}}\Xi^2(125)\Gamma(\bar{5}34)\overline{A}_{\bar{2}^{G^{-1}}}A_{\bar{3}}A_{\bar{4}}+\frac{\hat{5}}{\hat{1}}\Xi^2(154)\frac{\hat{2}}{\hat{5}}\Theta^2(\bar{5}23)\overline{A}_{\bar{2}^{G^{-1}}}A_{\bar{3}^{G^{-1}}}A_{\bar{4}}\\	\nonumber&+&\frac{\hat{5}}{\hat{1}}\Xi^2(154)\frac{\hat{3}}{\hat{5}}\Theta^3(\bar{5}23)A_{\bar{2}^{G^{-1}}}\overline{A}_{\bar{3}^{G^{-1}}}A_{\bar{4}}\\
\nonumber&-&\frac{\hat{2}}{\hat{1}}\Xi^2(125)\Gamma(\bar{5}34)\overline{A}_{\bar{2}}A_{\bar{3}^G}A_{\bar{4}^G}-\frac{\hat{5}}{\hat{1}}\Xi^2(154)\frac{\hat{2}}{\hat{5}}\Theta^1(\bar{5}23)\overline{A}_{\bar{2}}A_{\bar{3}}A_{\bar{4}^G}\\
\nonumber&-&\frac{\hat{5}}{\hat{1}}\Xi^2(154)\frac{\hat{3}}{\hat{5}}\Theta^3(\bar{5}23)A_{\bar{2}}\overline{A}_{\bar{3}}A_{\bar{4}^G}\\
\nonumber&-&\frac{\hat{5}}{\hat{1}}\Xi^3(125)\frac{\hat{3}}{\hat{5}}\Theta^2(\bar{5}34)A_{\bar{2}^G}\overline{A}_{\bar{3}}A_{\bar{4}}-\frac{\hat{5}}{\hat{1}}\Xi^3(125)\frac{\hat{4}}{\hat{5}}\Theta^3(\bar{5}34)A_{\bar{2}^G}A_{\bar{3}}\overline{A}_{\bar{4}}\\
\nonumber&-&\frac{\hat{3}}{\hat{1}}\Xi^3(154)\Gamma(\bar{5}23)A_{\bar{2}^G}A_{\bar{3}^G}\overline{A}_{\bar{4}}\\
\nonumber&+&\frac{\hat{5}}{\hat{1}}\Xi^3(125)\frac{\hat{3}}{\hat{5}}\Theta^2(\bar{5}34)A_{\bar{2}}\overline{A}_{\bar{3}^{G^{-1}}}A_{\bar{4}^{G^{-1}}}+\frac{\hat{5}}{\hat{1}}\Xi^3(125)\frac{\hat{4}}{\hat{5}}\Theta^3(\bar{5}34)A_{\bar{2}}A_{\bar{3}^{G^{-1}}}\overline{A}_{\bar{4}^{G^{-1}}}\\
\nonumber&+&\frac{\hat{4}}{\hat{1}}\Xi^3(154)\Gamma(\bar{5}23)A_{\bar{2}}A_{\bar{3}}\overline{A}_{\bar{4}^{G^{-1}}}\\
\nonumber&+&\frac{\hat{2}}{\hat{1}}\Xi^2(1234)\overline{A}_{\bar{2}^{G^{-1}}}A_{\bar{3}}A_{\bar{4}}-\frac{\hat{2}}{\hat{1}}\Xi^2(1234)\overline{A}_{\bar{2}}A_{\bar{3}^G}A_{\bar{4}}-\frac{\hat{2}}{\hat{1}}\Xi^2(1234)\overline{A}_{\bar{2}}A_{\bar{3}}A_{\bar{4}^G}\\
\nonumber\\
\nonumber&-&\frac{\hat{3}}{\hat{1}}\Xi^3(1234)A_{\bar{2}^G}\overline{A}_{\bar{3}}A_{\bar{4}}+\frac{\hat{3}}{\hat{1}}\Xi^3(1234)A_{\bar{2}}\overline{A}_{\bar{3}^{G^{-1}}}A_{\bar{4}}-\frac{\hat{3}}{\hat{1}}\Xi^3(1234)A_{\bar{2}}\overline{A}_{\bar{3}}A_{\bar{4}^G}\\
\nonumber\\
\nonumber&-&\frac{\hat{4}}{\hat{1}}\Xi^4(1234)A_{\bar{2}^G}A_{\bar{3}}\overline{A}_{\bar{4}}-\frac{\hat{4}}{\hat{1}}\Xi^4(1234)A_{\bar{2}}A_{\bar{3}^G}\overline{A}_{\bar{4}}+\frac{\hat{4}}{\hat{1}}\Xi^4(1234)A_{\bar{2}}A_{\bar{3}}\overline{A}_{\bar{4}^{G^{-1}}}\bigg\}+\cdots.
\end{eqnarray}
We shall persevere and collect terms and use relations (\ref{eq:Theta}) and (\ref{eq:Xi}),
\begin{eqnarray}	\nonumber\cdots-\epsilon\int_{234}&\bigg\{&\left(\frac{\hat{2}}{\hat{1}}\right)^2\Gamma(1234)\overline{A}_{\bar{2}^{G^{-1}}}A_{\bar{3}^{G^{-1}}}A_{\bar{4}^{G^{-1}}}-\left(\frac{\hat{3}}{\hat{1}}\right)^2\Gamma(1234)A_{\bar{2}^{G^{-1}}}\overline{A}_{\bar{3}^{G^{-1}}}A_{\bar{4}^{G^{-1}}}\\
\nonumber&-&\left(\frac{\hat{4}}{\hat{1}}\right)^2\Gamma(1234)A_{\bar{2}^{G^{-1}}}A_{\bar{3}^{G^{-1}}}\overline{A}_{\bar{4}^{G^{-1}}}-\left(\frac{\hat{2}}{\hat{1}}\right)^2\Upsilon(154)\Gamma(\bar{5}23)\overline{A}_{\bar{2}^{G^{-1}}}A_{\bar{3}^{G^{-1}}}A_{\bar{4}}\\
\nonumber&-&\left(\frac{\hat{3}}{\hat{1}}\right)^2\Upsilon(154)\Gamma(\bar{5}23)A_{\bar{2}^{G^{-1}}}\overline{A}_{\bar{3}^{G^{-1}}}A_{\bar{4}}+\left(\frac{\hat{4}}{\hat{1}}\right)^2\Upsilon(154)\Gamma(\bar{5}23)A_{\bar{2}^G}A_{\bar{3}^G}\overline{A}_{\bar{4}}\\
\nonumber&+&\left(\frac{\hat{2}}{\hat{1}}\right)^2\Upsilon(125)\Gamma(\bar{5}34)\overline{A}_{\bar{2}}A_{\bar{3}^G}A_{\bar{4}^G}-\left(\frac{\hat{3}}{\hat{1}}\right)^2\Upsilon(125)\Gamma(\bar{5}34)A_{\bar{2}}\overline{A}_{\bar{3}^{G^{-1}}}A_{\bar{4}^{G^{-1}}}\\
\nonumber&-&\left(\frac{\hat{4}}{\hat{1}}\right)^2\Upsilon(125)\Gamma(\bar{5}34)A_{\bar{2}}A_{\bar{3}^{G^{-1}}}\overline{A}_{\bar{4}^{G^{-1}}}+\left(\frac{\hat{2}}{\hat{1}}\right)^2\Upsilon(1234)\overline{A}_{\bar{2}}A_{\bar{3}^G}A_{\bar{4}}\\
\nonumber&-&\left(\frac{\hat{3}}{\hat{1}}\right)^2\Upsilon(1234)A_{\bar{2}}\overline{A}_{\bar{3}^{G^{-1}}}A_{\bar{4}}+\left(\frac{\hat{4}}{\hat{1}}\right)^2\Upsilon(1234)A_{\bar{2}}A_{\bar{3}^G}\overline{A}_{\bar{4}}\\
\nonumber&-&\left(\frac{\hat{2}}{\hat{1}}\right)^2\left\{\Upsilon(152)\Gamma(\bar{5}34)+\Upsilon(1234)\right\}\overline{A}_{\bar{2}^{G^{-1}}}A_{\bar{3}}A_{\bar{4}}\\
\nonumber&+&\left(\frac{\hat{3}}{\hat{1}}\right)^2\left\{\Upsilon(152)\Gamma(\bar{5}34)+\Upsilon(1234)\right\}A_{\bar{2}^G}\overline{A}_{\bar{3}}A_{\bar{4}}\\
\nonumber&+&\left(\frac{\hat{4}}{\hat{1}}\right)^2\left\{\Upsilon(154)\Gamma(\bar{5}23)+\Upsilon(1234)\right\}A_{\bar{2}^G}A_{\bar{3}}\overline{A}_{\bar{4}}\\
\nonumber&+&\left(\frac{\hat{2}}{\hat{1}}\right)^2\left\{\Upsilon(152)\Gamma(\bar{5}34)+\Upsilon(1234)\right\}\overline{A}_{\bar{2}}A_{\bar{3}}A_{\bar{4}^G}\\
\nonumber&+&\left(\frac{\hat{3}}{\hat{1}}\right)^2\left\{\Upsilon(154)\Gamma(\bar{5}23)+\Upsilon(1234)\right\}A_{\bar{2}}\overline{A}_{\bar{3}}A_{\bar{4}^G}\\
\nonumber&-&\left(\frac{\hat{4}}{\hat{1}}\right)^2\left\{\Upsilon(154)\Gamma(\bar{5}23)+\Upsilon(1234)\right\}A_{\bar{2}}A_{\bar{3}}\overline{A}_{\bar{4}^{G^{-1}}}\bigg\}+\cdots.
\end{eqnarray}
Fortunately, now, we have already calculated the above expressions enclosed in paranthesis in  $\Gamma$ and $\Upsilon$ in the previous calculation of $\delta A$ in appendix (\ref{ap:A}) so we do not need to do these again. We can reach the result
\begin{eqnarray}
\label{eq:dbarA3rdorder}	\nonumber\cdots-\epsilon\int_{234}&\bigg\{&-\left(\frac{\hat{2}}{\hat{1}}\right)^2\frac{\hat{1}\hat{q}\overline{A}_{\bar{2}^{G^{-1}}}A_{\bar{3}^{G^{-1}}}A_{\bar{4}^{G^{-1}}}}{(q,2)(q,2+3)}-\left(\frac{\hat{3}}{\hat{1}}\right)^2\frac{\hat{1}\hat{q}A_{\bar{2}^{G^{-1}}}\overline{A}_{\bar{3}^{G^{-1}}}A_{\bar{4}^{G^{-1}}}}{(q,2)(q,2+3)}-\\
\nonumber&-&\left(\frac{\hat{4}}{\hat{1}}\right)^2\frac{\hat{1}\hat{q}A_{\bar{2}^{G^{-1}}}A_{\bar{3}^{G^{-1}}}\overline{A}_{\bar{4}^{G^{-1}}}}{(q,2)(q,2+3)}-\left(\frac{\hat{2}}{\hat{1}}\right)^2\frac{\hat{1}\hat{q}\overline{A}_{\bar{2}^{G^{-1}}}A_{\bar{3}^{G^{-1}}}A_{\bar{4}}}{(q,2)(q,4)}-\\
\nonumber&-&\left(\frac{\hat{3}}{\hat{1}}\right)^2\frac{\hat{1}\hat{q}A_{\bar{2}^{G^{-1}}}\overline{A}_{\bar{3}^{G^{-1}}}A_{\bar{4}}}{(q,2)(q,4)}+\left(\frac{\hat{4}}{\hat{1}}\right)^2\frac{\hat{1}\hat{q}A_{\bar{2}^G}A_{\bar{3}^G}\overline{A}_{\bar{4}}}{(q,2)(q,4)}+\\
\nonumber&+&\left(\frac{\hat{2}}{\hat{1}}\right)^2\frac{\hat{1}\hat{q}\overline{A}_{\bar{2}}A_{\bar{3}^G}A_{\bar{4}^G}}{(q,3)(q,1)}-\left(\frac{\hat{3}}{\hat{1}}\right)^2\frac{\hat{1}\hat{2}A_{\bar{2}}\overline{A}_{\bar{3}^{G^{-1}}}A_{\bar{4}^{G^{-1}}}}{(q,3)(q,1)}-\\
\nonumber&-&\left(\frac{\hat{4}}{\hat{1}}\right)^2\frac{\hat{1}\hat{q}A_{\bar{2}}A_{\bar{3}^{G^{-1}}}\overline{A}_{\bar{4}^{G^{-1}}}}{(q,3)(q,1)}+\left(\frac{\hat{2}}{\hat{1}}\right)^2\frac{\hat{1}\hat{q}\overline{A}_{\bar{2}}A_{\bar{3}^G}A_{\bar{4}}}{(q,4)(q,4+1)}-\\
\nonumber&-&\left(\frac{\hat{3}}{\hat{1}}\right)^2\frac{\hat{1}\hat{q}A_{\bar{2}}\overline{A}_{\bar{3}^{G^{-1}}}A_{\bar{4}}}{(q,4)(q,4+1)}+\left(\frac{\hat{4}}{\hat{1}}\right)^2\frac{\hat{1}\hat{q}A_{\bar{2}}A_{\bar{3}^G}\overline{A}_{\bar{4}}}{(q,4)(q,4+1)}-\\
\nonumber&-&\left(\frac{\hat{2}}{\hat{1}}\right)^2\frac{\hat{1}\hat{q}\overline{A}_{\bar{2}^{G^{-1}}}A_{\bar{3}}A_{\bar{4}}}{(q,3)(q,3+4)}+\left(\frac{\hat{3}}{\hat{1}}\right)^2\frac{\hat{1}\hat{q}A_{\bar{2}^G}\overline{A}_{\bar{3}}A_{\bar{4}}}{(q,3)(q,3+4)}-\\
\nonumber&+&\left(\frac{\hat{4}}{\hat{1}}\right)^2\frac{\hat{1}\hat{q}A_{\bar{2}^G}A_{\bar{3}}\overline{A}_{\bar{4}}}{(q,3)(q,3+4)}+\left(\frac{\hat{2}}{\hat{1}}\right)^2\frac{\hat{1}\hat{q}\overline{A}_{\bar{2}}A_{\bar{3}}A_{\bar{4}^G}}{(q,1)(q,1+2)}+\\
\nonumber&+&\left(\frac{\hat{3}}{\hat{1}}\right)^2\frac{\hat{1}\hat{q}A_{\bar{2}}\overline{A}_{\bar{3}}A_{\bar{4}^G}}{(q,1)(q,1+2)}-\left(\frac{\hat{4}}{\hat{1}}\right)^2\frac{\hat{1}\hat{q}A_{\bar{2}}A_{\bar{3}}\overline{A}_{\bar{4}^{G^{-1}}}}{(q,1)(q,1+2)}\bigg\}+\cdots.\\
\end{eqnarray}
\bibliography{tau} 
\bibliographystyle{jhep}
\end{document}